%% file: prePrint-plain.tex
\documentclass[12pt]{article}

\usepackage[utf8]{inputenc}
\usepackage{natbib}
\usepackage{paralist}
\usepackage{amsmath}
\usepackage{amsfonts}
\usepackage{stix}
\usepackage{url}
\usepackage{mathtools}
\usepackage{color}
\usepackage{xcolor}
\usepackage{cprotect}
\usepackage{xspace}
\usepackage{amsthm}
\usepackage{multirow}
\usepackage{clrscode}
\usepackage{fancyhdr}
\usepackage{textcomp}
\usepackage{listings}
\usepackage{makeidx}
\usepackage{ifthen}
\usepackage{hyperref}

\usepackage{nicefrac}
\usepackage{algorithmic}
\usepackage{algorithm}
\usepackage{calc}
\usepackage[Lenny]{fncychap}
\usepackage{bm}
\usepackage{graphicx}
\usepackage{tabularx}
\usepackage{subfig}

\usepackage{pgfplots}
\usepgfplotslibrary{polar}
\usepackage{pgfplotstable, booktabs}
\usepackage{makecell}
\pgfplotsset{compat=1.9}
\usetikzlibrary{shapes}

\usepackage{chngcntr}

\usepackage[margin=35mm]{geometry} 

\title{On floating point precision in computational fluid dynamics using OpenFOAM}

\author{\small F. Brogi$^1$, S. Bn\`a$^2$, G. Boga$^{2,3}$, G. Amati$^2$, T. Esposti Ongaro$^1$, M. Cerminara$^1$}

\date{ \small
	$^1$Istituto Nazionale di Geofisica e Vulcanologia, Pisa, Italy \\%
	$^2$CINECA Supercomputing Centre, Via Magnanelli 6/3, 40033 Casalecchio di Reno, Italy \\ %
	$^3$ DIEF, University of Modena and Reggio Emilia, 41125 Modena, Italy \\
}

\begin{document}
\maketitle
\input{00-abstract}

\input{0-introduction}

\input{1-solution_quality}
\input{2-performance_gain}
\input{3-scaling_analysis}

\input{4-GPU}

\input{5-conclusions}
\input{6-appendix}

\input{7-Aknwoledgments}

\clearpage
\bibliography{references}

\end{document}

%% file: 00-abstract.tex
\begin{abstract}
Thanks to the computational power of modern cluster machines, numerical simulations can provide, with an unprecedented level of details, new insights into fluid mechanics. However, taking full advantage of this hardware remains challenging since data communication remains a significant bottleneck to reaching peak performances.
Reducing floating point precision is a simple and effective way to reduce data movement and improve the computational speed of most applications. Nevertheless, special care needs to be taken to ensure the quality and convergence of computed solutions, especially when dealing with complex fluid simulations. 
In this work, we analyse the impact of reduced (single and mixed compared to double) precision on computational performance and accuracy for computational fluid dynamics. 
Using the open source library OpenFOAM, we consider incompressible, compressible, and multiphase fluid solvers for testing on relevant benchmarks for flows in the laminar and turbulent regime and in the presence of shock waves. Computational gain and changes in the scalability of applications in reduced precision are also discussed. In particular, an ad hoc theoretical model for the strong scaling allows us to interpret and understand the observed behaviors, as a function of floating point precision and hardware specifics. Finally, we show how reduced precision can significantly speed up a hybrid CPU-GPU implementation, made available to OpenFOAM end-users recently, that simply relies on a GPU linear algebra solver developed by hardware vendors.

\end{abstract}


%% file: 0-introduction.tex
\section{Introduction\label{sec:introduction}}

An increasingly wider community is choosing the open source software \citet{openfoam.com,openfoam.org} as a flexible tool to perform numerical simulation in continuum mechanics including fluid dynamics, solid mechanics and electromagnetics. OpenFOAM's modular structure allows end users to easily build new solvers and developers to add new features, constantly enlarging the range of possible applications of interest for both the academy and industry \citep{Dipaolo2021,Rauter2021}. However, made exception for a recent coordinated effort (www.exafoam.eu), relatively less attention have been paid by OpenFOAM developers to computational performances. In particular, aspects such as its parallel efficiency on massively parallel machines remain challenging \citep{Axtmann2016}.
Individual research groups have managed to resolve implementation bottlenecks and improve performance of OpenFOAM,  (see \citet{Bna2020} and references therein) but, often the lack of generality of their implementations have prevented their direct inclusion in the official OpenFOAM distributions.
 
Tailored optimizations strategies are indeed required for any application to exploit the computational power of current and next-generation supercomputing systems. Modern HPC machines typically achieve their full computational power using heterogeneous hardware (e.g. CPU-GPU) and a huge number (up to millions) of computing units. Optimizing and reducing data movement, including memory access and I/O, is therefore a basic requirements to reach peak performances. The strong imbalance between the computational power and memory bandwidth still remains a major issue and it has not been attenuated by the latest technological trends. Performing arithmetic operations remains indeed several orders of magnitude faster than accessing data in memory, or performing communications between different computational nodes of a cluster machine. From this perspective, most of real applications, including OpenFOAM, are and will become more and more memory-bound \citep{Abdelfattah2021,Succi2019}.

Changing algorithms can help to significantly improve the communication to computation ratio (i.e. the operational intensity) of an application but it may be a challenging and time consuming exercise, especially for complex general-purpose codes such as OpenFOAM. 
Reducing floating point precision arithmetic is instead often a simple but effective way of cutting down data movement and increasing the computational speed for most applications. 
On modern CPUs, performing operations in single precision (32-bit) is twice as fast as in double precision (64-bit), since the amount of data moved in memory is halved and arithmetic operations are twice as fast as double precision \citep{Baboulin2009}.
When considering less conventional hardware such as NVIDIA GPU Tensor Cores the computational gain becomes even more attractive. These architectures in fact support half precision arithmetic with dedicated functional units (accelerators) in the hardware that makes low precision much faster than higher precision \citep{Haidar2018}. With this hardware, half precision arithmetic has a theoretical speedup of $16\times$ instead of the expected  $4\times$, with respect to double precision \citep{Abdelfattah2019}. 
 However, using low floating point precision is not always possible. Traditionally, CFD codes work with double precision since for complex fluid problems linear algebra solvers may not converge or not provide the solution with the required degree of accuracy. A number of studies have proposed the use of mixed precision algorithms to unleash the power of multi-precision hardware without sacrificing accuracy or numerical stability (\citet{Abdelfattah2021} for a review).

In OpenFOAM, a mixed precision feature has been released with version v1906 and is implemented in its complex framework following a rather simple idea: all the code is compiled in single precision except for the linear algebra solvers, which work in double precision. The reduced memory consumption alleviates the bandwidth bottleneck of OpenFOAM, thus increasing the computational speed of almost any application. However, to our knowledge, there are no systematic studies on the use of reduced precision computations in CFD applications in the literature, making an exception for more specific case studies such as the recent work on fluid the lattice Boltzmann method based fluid solvers \citep{Lehmann2022}. 

In this work, we analyse and discuss the impact of floating point precision reduction (single and mixed with respect to double) on real CFD applications using OpenFOAM. In particular, we consider two important aspects such as the convergence and accuracy of computed solutions, the computational performance on both CPU and hybrid CPU-GPU hardware as well as the parallel efficiency of CFD applications. With the aid of theoretical and experimental analysis, we describe how precision reduction affects the individual parts of the applications that are commonly present in CFD solvers. In order to try to keep our results as much as possible of general interest, both incompressible and compressible solvers have been selected for testing, since they may represent the basis for any more complex solver to be built on (e.g. multiphase solvers). The quality of computed solutions (accuracy and convergence) of these solvers are tested considering important flow phenomena in laminar and turbulent flow regimes as well as the presence of shock wave discontinuities. Performance gain and change in the scaling behaviour of applications on parallel machines due to precision reduction are also considered. A theoretical model for the strong scaling of applications is also developed, which allows us to explain and better understand the changes observed with reduced floating point precision and different hardware specifics. Finally, we demonstrate how significant can be the effect of mixed precision on computational performance, with a speedup of $2.4\times$, when considering the hybrid CPU-GPU implementation of OpenFOAM that has been recently made available to users.

%% file: 1-solution_quality.tex
\section{Impact of precision reduction on the quality of computed solutions}
We here analyse the impact of reduced floating point precision (single and mixed) on two important aspects of CFD applications: the accuracy and convergence of computed solutions.
Given the large number of CFD applications available in OpenFOAM, we limit our study to solvers for single phase incompressible and compressible flows. It is well known in fact that these solvers face different computational challenges and represent the ground for almost any more complex solver (e.g. multiphase). Here, and in the rest of the paper, we use OpenFOAM v1912 for our tests, if not otherwise stated. The classical \texttt{icoFOAM} (incompressible) and \texttt{rhoPimpleFOAM} (compressible) solvers have been selected to solve radically different benchmarks: the laminar 3D lid-driven cavity, the decay of isotropic turbulence, the shock tube and the starting compressible jet.
 The lid driven cavity is a well known benchmark for incompressible solvers \citep{Shankar2000}, especially when considering the laminar steady state regime. The isotropic turbulence benchmark \citep{Pirozzoli2004} is also a standard test and it is well suited to evaluate the effect of reduced precision on numerical simulations with turbulence, here in the weakly compressible regime. The Sod shock tube test \citep{Sod1978} is instead used to understand whether discontinuous solutions are affected by floating point precision representation. The fully compressible dynamic of a starting jet (high Mach and high Reynolds number) is used to test the behaviour of the numerical solver with reduced precision for transient compressible problems. Finally, as an example of a complex real-case application, we consider the simulation of a compressible, multiphase turbulent volcanic plume with the OpenFOAM-based solver \texttt{ASHEE} \citep{Cerminara2016}.

\subsection{Lid-driven cavity}
In this benchmark, an incompressible fluid is initially at rest in a cubic cavity. A tangential velocity is imposed on the top boundary and no-slip conditions (wall) are enforced on all the other boundaries.
The flow regime in the cavity depends only on the dimensionless Reynolds number ($\mathrm{Re}=UL/\nu$, with $U$, the velocity of the moving lid, L the length of the cavity and $\nu$ the kinematic viscosity). The interested reader may refer to \citet{Shankar2000} for a complete review. Here we consider the laminar steady state regime for two different Reynolds numbers ($\mathrm{Re} = 100, 1000$). For this benchmark, we used the incompressible \texttt{icoFOAM} (v1912) solver compiled in single, mixed and double precision.

\begin{figure}[ht]
 \centering
\subfloat[][\emph{$\mathrm{Re}=100$}.]
{\includegraphics[width=0.5\columnwidth]{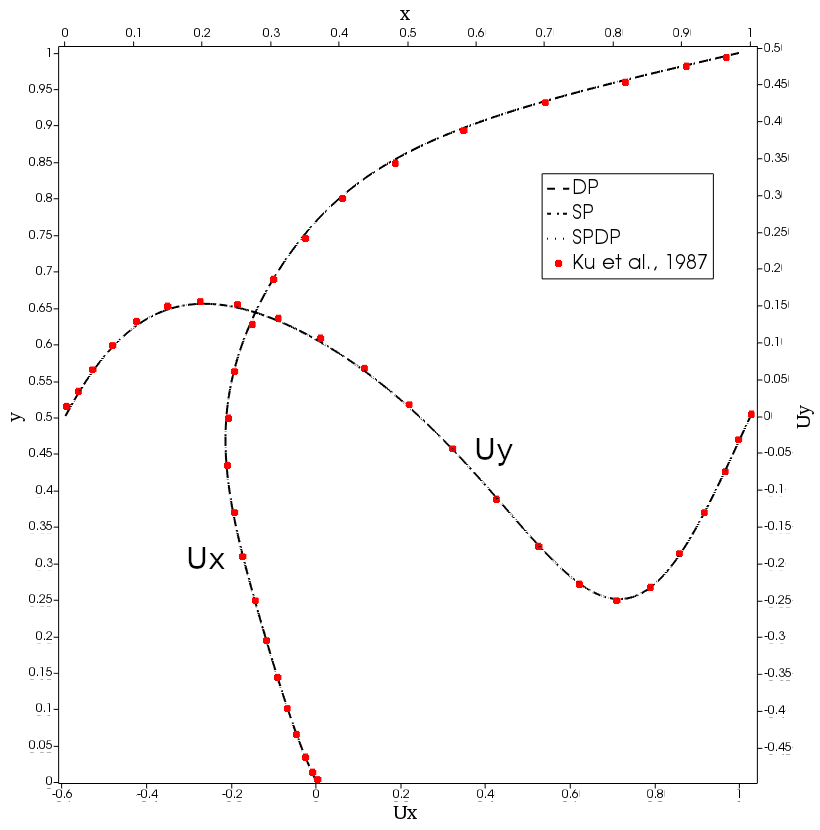}}
\subfloat[][\emph{$\mathrm{Re}=1000$}.]
{\includegraphics[width=0.5\columnwidth]{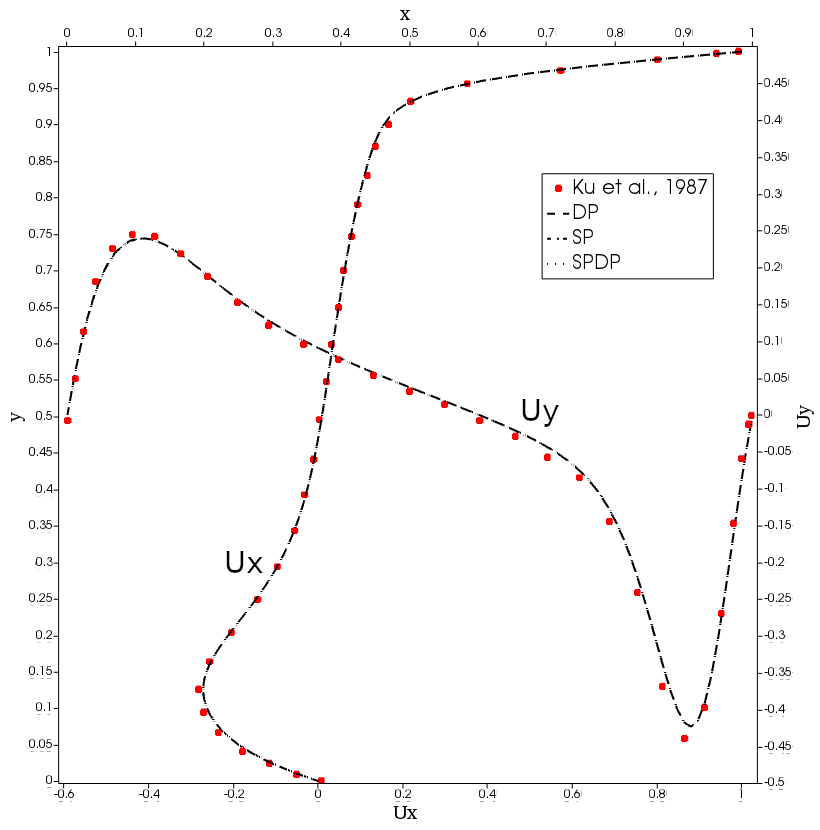}}
\caption{Lid-driven cavity benchmark solved with \texttt{icoFOAM} solver, using a mesh size of $100^3$ cells, in single (SP), double (DP) and mixed precision (SPDP). Velocity profiles of $U_x$ along y-axis centerline (left vertical axis) and $U_y$ along x-axis centerline (top horizontal axis) are compared with those extracted from the figures of \citet{Ku1987}.}
\label{fig:cavityVelProfiles}
\end{figure}

 All the simulations converge to physical solutions (Fig. \ref{fig:cavityVelProfiles}) and regardless of the precision, they are consistent with a well known reference study \citep{Ku1987}. In particular, our results demonstrate that the single precision is sufficient for the fluid solver to provide accurate velocity profiles along the cavity centerlines (fig. \ref{fig:cavityVelProfiles}). 

\subsection{Sod shock tube}

\begin{figure}[ht]
 \centering
 \includegraphics[width=0.8\columnwidth]{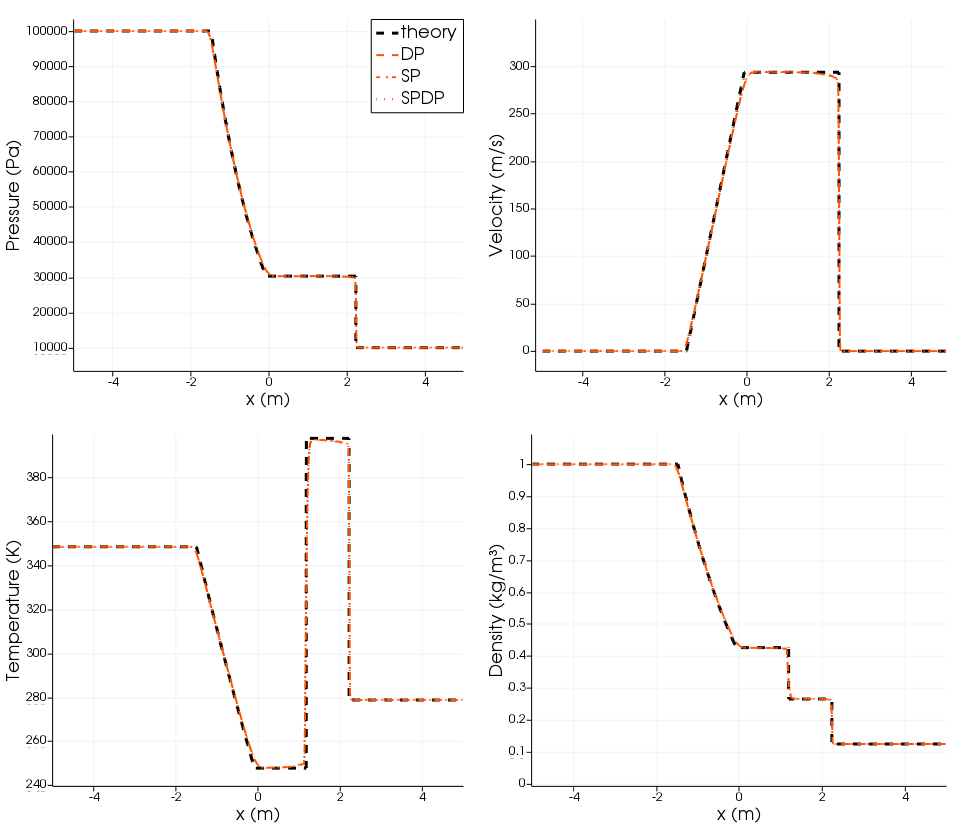}
 \caption{Sod shock tube benchmark solved with \texttt{rhoPimpleFOAM} solver in single (SP), double (DP) and mixed precision (SPDP). Pressure, temperature, velocity and density profiles (orange) are compared with the theoretical ones (black).}
\label{fig:shockTube}
 \end{figure}

This test considers the dynamic that develops inside a 1D tube when the diaphragm, separating high-pressure and low-pressure regions of a gas, is removed. Typically, a rarefaction wave propagates towards the high-pressure region and a shock wave (a discontinuity on all fluid variables) forms and moves through the low-pressure region. Across these discontinuities numerical schemes, that requires the solution to be smooth for derivatives to not diverge, normally face accuracy and stability issues. 
The \texttt{rhoPimpleFOAM} solver (v1912) compiled in single, double and mixed precision is here used to produce the numerical solution. The initial conditions are reported in the caption of Figure \ref{fig:shockTube}. The tube length is resolved with 1000 cells and the Courant number is set to $\mathrm{Co}=0.01$ in all simulations since for this value the double precision built is proven to be accurate. 
As in the previous benchmark, despite the precision used, the solver converges to a physical solution that is in remarkable good agreement with the theoretical one \citep{Sod1978}. The presence of contact and shock discontinuities are well resolved and important numerical artefacts (e.g. spurious oscillations) are not evident in the solutions with reduced precision. Setting larger values of the Courant number $>0.01$ causes the numerical solution to become rapidly less accurate. However, in this case, reduced precision does not change or decrease the quality of the numerical solution appreciably. 

\subsection{Compressible decaying homogeneous and isotropic turbulence}
\label{sec:isotropicTurbulence}
Turbulence is one of the main non-linear complexities of fluid dynamics. The problem arises when the dissipation scale is much smaller than the integral scale ($L$) of the flow and the laminar solution of the Navier-Stokes equations becomes unstable: eddies start to form and break until the dissipation scale is reached. This is the Kolmogorov micro-scale $\eta$. The ratio between the integral and the dissipation scale grows with the Reynolds number $L/\eta\propto\mathrm{Re}^\frac{3}{4}$. A simulation that is able to capture all the scales (from the largest integral one, down to the smallest dissipation scale) is called Direct Numerical Simulations (DNS). When the flow is far from boundaries, the typical behaviour of turbulence is described by the energy cascade of the kinetic energy spectrum. When it is impossible to resolve all the relevant scales, one of the possibilities is to use the Large Eddy Simulations (LES) method, to take into account of the energy cascade in the sub-grid terms. A general description of the physical, mathematical and numerical problem can be found in \citep{blaisdell1991numerical, Wang1993, Garnier1999, pope2000, Honein2004, Pirozzoli2004, lesieur2005, Liao2009, Bernardini2009, bernardini_2014, Bernardini2021} and in references therein. Homogeneous and isotropic decaying turbulence is a classic benchmark used to test the capabilities of numerical codes to solve turbulence far from boundaries. Here, we are following the same procedure described by \citet{Cerminara2016,Cerminara2016phd}, by using the \texttt{ASHEE} code in its single-phase and mixed-precision configuration. The objective is to compare the mixed-precision solution with the double-precision one, having already validated the latter with the eight-order scheme by~\citet{Pirozzoli2004}. Double precision is often needed in problems facing turbulence, to solve as accurately as possible the non-linear advection terms present in the Navier-Stokes equations. Indeed, these terms tend to grow with the spatial scale, becoming larger than the (mainly linear) dissipation terms. For this reason, double precision becomes even more important in LES, where the smallest scale is much larger than the dissipation scale and its contribution is taken into account using non-linear sub-grid scale turbulence viscosity terms \citep{lesieur2005}.

\begin{figure}[ht]
 \centering
 \includegraphics[width=0.49\columnwidth]{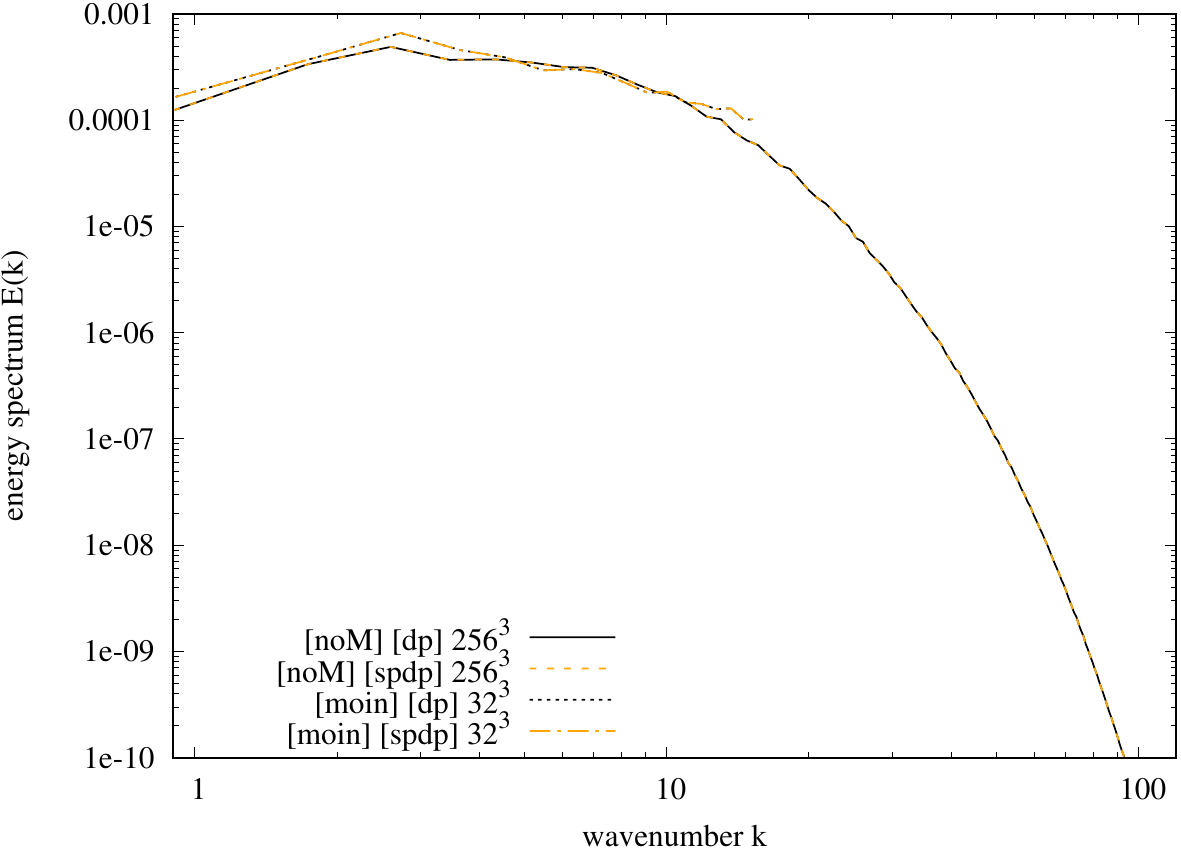}
 \includegraphics[width=0.49\columnwidth]{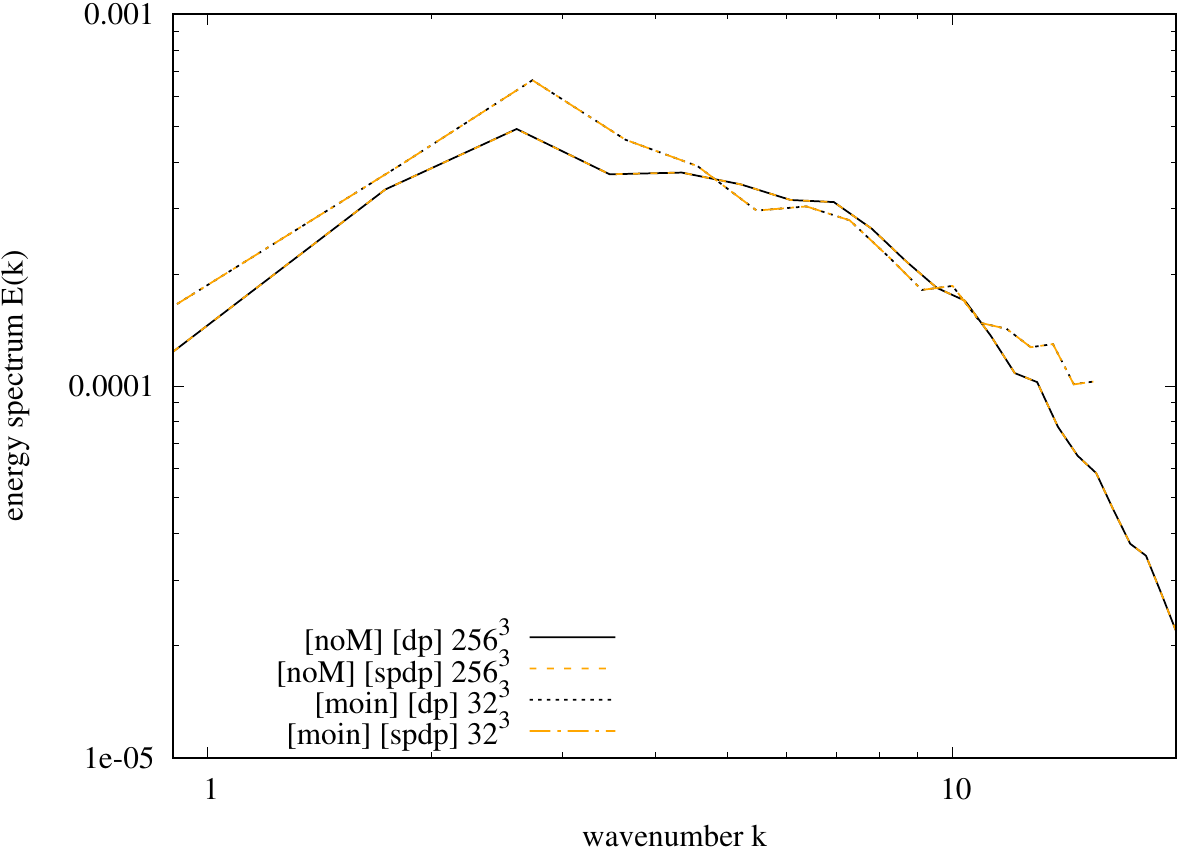}
 \caption{Kinetic energy spectrum of the DNS (noM) and of the LES (moin) in double precision (dp) and mixed precision (spdp), after $t=5.5\tau_\mathrm{e}$. The right panel is just a zoom in the region resolved by the LES.}
\label{fig:hit_energySpectrum}
 \end{figure}

The DNS test is performed using a mesh size of $256^3$ cells in a cubic box with side $L=2\pi$ and periodic boundary conditions. In this way, boundary effects can be neglected. The initial spectrum is the same described in Section 5.2 of \citet{Cerminara2016phd}, so that the root-mean-square Mach number is $\mathrm{Ma}_\mathrm{rms}=0.2$, the initial Taylor microscale is $\lambda=0.5$, the eddy turnover time is $\tau_\mathrm{e}=3.66$, the Reynolds number based on the Taylor micro-scale is $\mathrm{Re}_\lambda = 116$, and the maximum wave-number is $k_\mathrm{max}=127$. In this way, its product with the Kolmogorov micro-scale $\eta = 0.023$ is large enough to have a proper DNS. The same initial condition is mapped into a $32^3$ mesh to perform a LES to be compared with the previous DNS results. The comparison has been performed by using both mixed and double precision. LES are executed by using the Moin's model \citep{Cerminara2016, Cerminara2016phd}. Simulations in single precision turned out to be unstable in this case. In Fig.~\ref{fig:hit_energySpectrum}, the spectrum of the kinetic energy obtained for all the 4 simulations is shown. Mixed precision works pretty well for this test case, with minor influence on the solution: the relative error with respect to the double precision case is $6.0\cdot10^{-7}$, and $7.2\cdot10^{-7}$, for the DNS and LES, respectively. 90\% of the error is contained in the region $k < 10$, and $k < 5$, respectively for DNS and LES. By using a different LES sub-grid model, the effect of precision can have a larger impact. For example, by using the dynamic WALE model \citep{Cerminara2016,Cerminara2016phd}, the relative discrepancy increases to $1.4\cdot10^{-4}$.

\subsection{Starting compressible square jet}

\begin{figure}[ht]
\centering
\includegraphics[width=0.7\columnwidth]{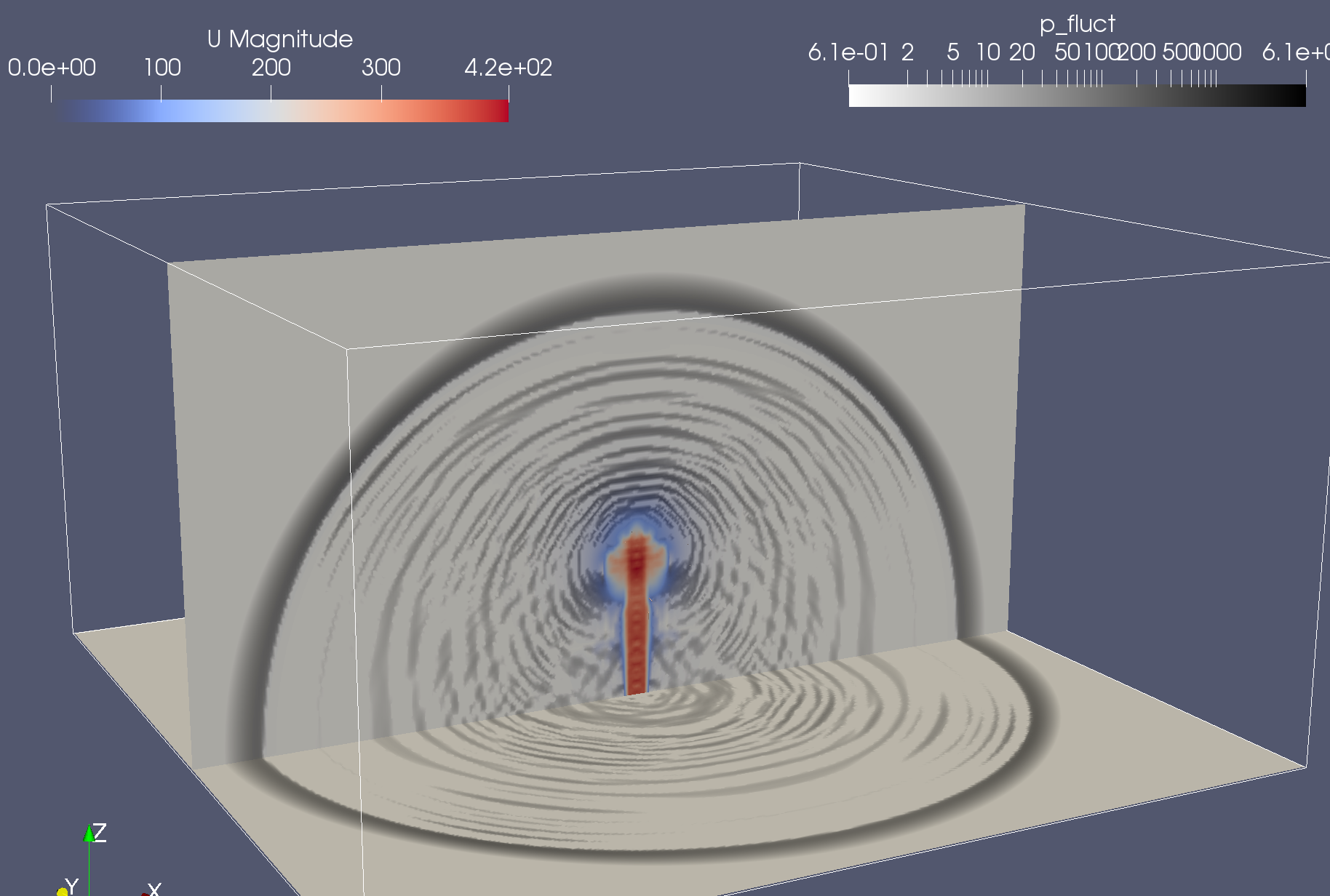}
\caption{Screenshot of LES simulation of the supersonic starting jet using \texttt{rhoPimpleFOAM} solver. The magnitude of the velocity field (in color) and pressure fluctuations $\delta{p}=p-p_{ref}$ (black and white in log scale) are shown.}
\label{fig:miniappScreenshot}
\end{figure}

A high-speed injection of a warm gas (330 K) into a static atmosphere with a slightly lower temperature (300 K) is considered. For simplicity, the jet fluid enters the atmospheric box through a square inlet (0.0635 m) resolved with a homogeneous structured mesh (no grid stretching). Given the low gas viscosity ($1.8\times10^{-5}$ Pa s) and high velocity (364 m/s) of the injected fluid, the jet flow is unsteady and compressible, characterised by high Mach number (>1 locally) and Reynolds number. Non-linear instabilities in the jet shear layer (such as Kelvin-Helmholtz instabilities), eddies, and pressure waves contribute to the formation of a turbulent buoyant plume in these conditions. As the jet enters the atmospheric box it forms a vortex and pushes the static air outwards, producing an intense pressure wave that propagates radially (Fig. \ref{fig:miniappScreenshot}). In our setup, the simulation is stopped when the first transient reaches the boundary of the computational domain and hence before the turbulent plume develops.
The solver \texttt{rhoPimpleFOAM} for compressible laminar and turbulent flow is used for this test case.
An LES approach combined with a one-equation eddy viscosity model \citep{Yoshizawa1986} is used to allow the solver to deal with the unresolved scales of the turbulent flow.
For testing, velocity and pressure probes are placed along the jet centerline, in the jet shear layer and far from the flow field in the atmosphere to record pressure waves.

\begin{figure}[ht]
 \centering
\subfloat[][\emph{}]
{\includegraphics[width=0.5\columnwidth]{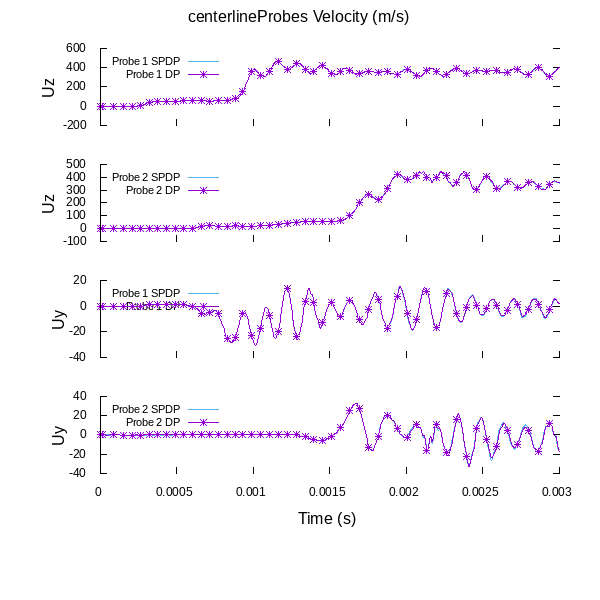}}
\subfloat[][\emph{}]
{\includegraphics[width=0.5\columnwidth]{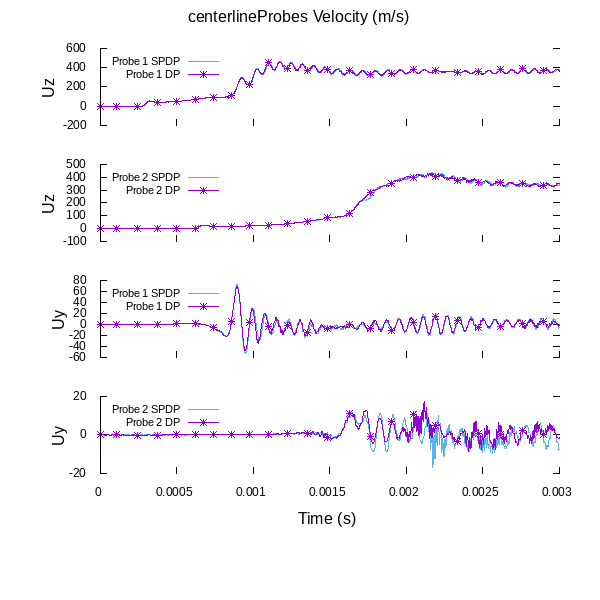}}
\caption{Times series of axial (Uz) and radial (Uy) components of the velocity field as recorded by two probes placed in the computational domain along the jet center-line for: (a) coarse mesh (16384000 cells); (b) refined mesh (131072000 cells).}
\label{fig:jetProbes}
\end{figure}

Single precision for this test case is not sufficient for linear algebra to converge. Therefore, no numerical solution is available with this precision. Mixed and double precision runs instead converge to physical solutions that result to be very similar. In particular, for a test case with a coarse mesh, the time series of pressure and velocity for all probes overlap almost completely (fig. \ref{fig:jetProbes}). When considering a test case with a refined mesh (double number of cells in the jet diameter),  double precision and mixed precision start to be different. The differences are particularly evident in those parts of the recorded signals where high-frequency content is present on all fluid variables (e.g. $Uy$ in fig. \ref{fig:jetProbes}b). The system sensitivity to tiny differences in the initial conditions may explain the changes observed in the numerical solutions with reduced precision and refined mesh. The impact of mixed precision on the accuracy of computed solutions can be evaluated by considering the statistical quantities of the fluid variables only (see section \ref{sec:isotropicTurbulence}).  

\subsection{Mixed precision in a real use-case: the simulation of a volcanic plume}
Moving to a geophysical scale problem, we present a numerical simulation of the evolution of an explosive volcanic eruption in a still atmosphere. Volcanic plumes are characterized by high Reynolds and Mach numbers. The multiphase gas-particle mixture injected into the atmosphere by these kinds of fascinating and catastrophic events is typically very hot (above 1000~K) and denser than the surrounding air (above 3~kg/m$^3$). Shocks, turbulence and acoustic fluctuations start to develop immediately after the beginning of the eruption. The plume initially rises because of its initial momentum. Then, turbulent mixing decreases its density due to atmospheric air entrainment and expansion and the plume may reach a level where the buoyancy starts to be positive (buoyancy reversal). At this level, the column starts to behave as a proper plume, accelerating upwards due to its buoyancy. From the neutral buoyancy level, the updrafting mixture decelerates, to finally spread laterally into the umbrella cloud. More details on the phenomenon can be found e.g. in \cite{Woods2010, Cerminara2015les, Cerminara2016phd, Neri2022}.
We now consider a test case to be solved with the OpenFOAM-based solver \texttt{ASHEE} \citep{Cerminara2016}. The \texttt{ASHEE} (ASH Equilibrium Eulerian) model, based on the dynamic LES model and an asymptotic expansion strategy of the full non-equilibrium multiphase Eulerian model, solves gas-polydisperse particle turbulent flows that characterize volcanic plumes. Although this approach is valid for dilute concentrations (volume fraction smaller than 1\%) of ash, larger particles can be also included in \texttt{ASHEE} using a Lagrangian approach with a two-way coupling regime.
For testing, we consider that the fragmented magma is injected into a still atmosphere from a 500~m wide inlet, with an initial velocity, temperature, and gas mass fraction equal to 236~m/s, 1050~K, and 5~wt.\%, respectively. This results in a mass flow rate equal to $2\times 10^8$~kg/s, which is representative of a Plinian eruption. Atmospheric stratification is modelled by using the U.S. Standard Atmosphere. The computational domain is 50 km high and extends along the 2 horizontal directions for 100~km. It is discretised using 40 million cells with an orthogonal mesh with constant grading both in the vertical and horizontal directions. 
The temporal discretization is based on the second-order Crank-Nicolson scheme, with an adaptive time stepping based on the Courant number ($\mathrm{Co} \leq 0.2$).
Numerical schemes and boundary conditions are described in detail in \citep{Cerminara2016, Cerminara2015les, Cerminara2016phd}. The bottom of the domain is treated as a thermally insulated slip wall, with a sink condition based on particle settling velocity to allow pyroclasts to deposit. The atmosphere is treated as an open boundary with input-output Dirichlet-Neumann conditions based on the velocity direction and a total pressure condition. The inlet has a prescribed velocity hyperbolic tangent profile to mimic conduit boundary effects. All these characteristics have been implemented in \texttt{ASHEE} by using the OpenFOAM infrastructure \citep{openfoam.com}.

\begin{figure}[ht]
 \centering
\subfloat[][\emph{}]
{\includegraphics[width=0.4\columnwidth]{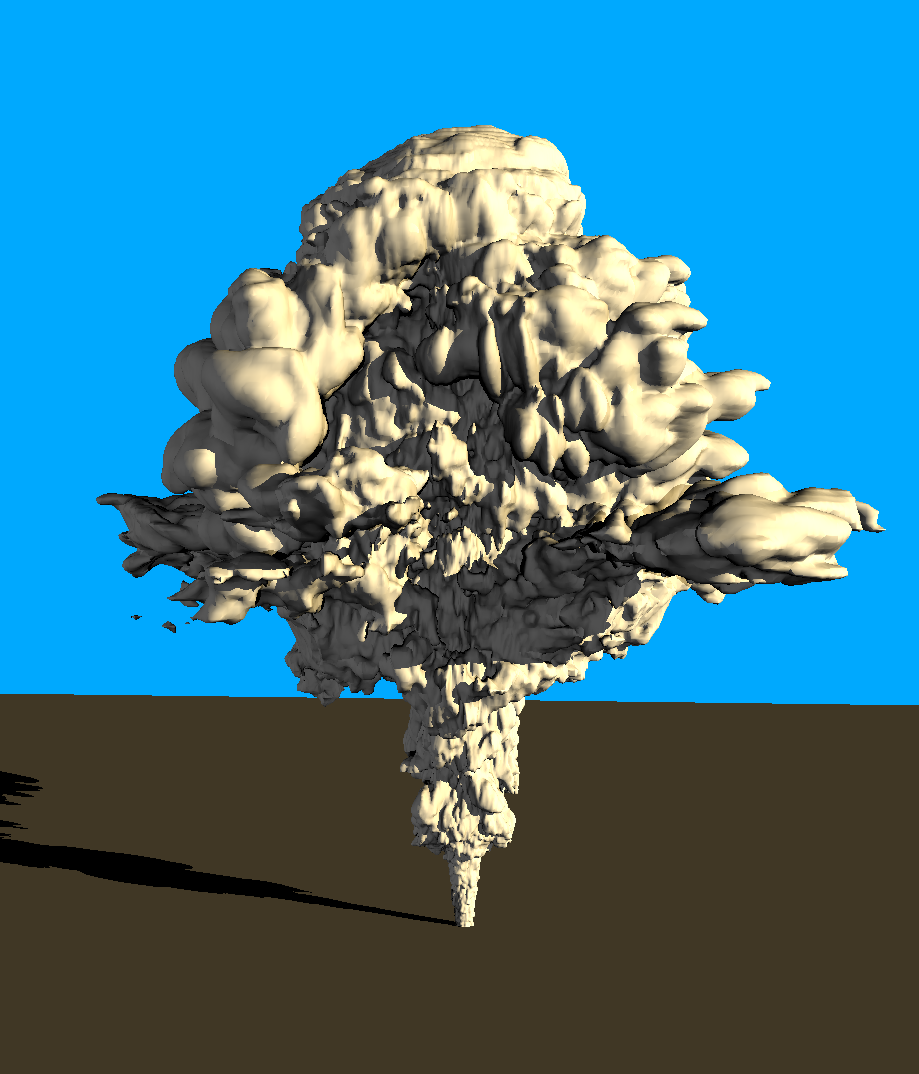}}
\subfloat[][\emph{}]
{\includegraphics[width=0.6\columnwidth]{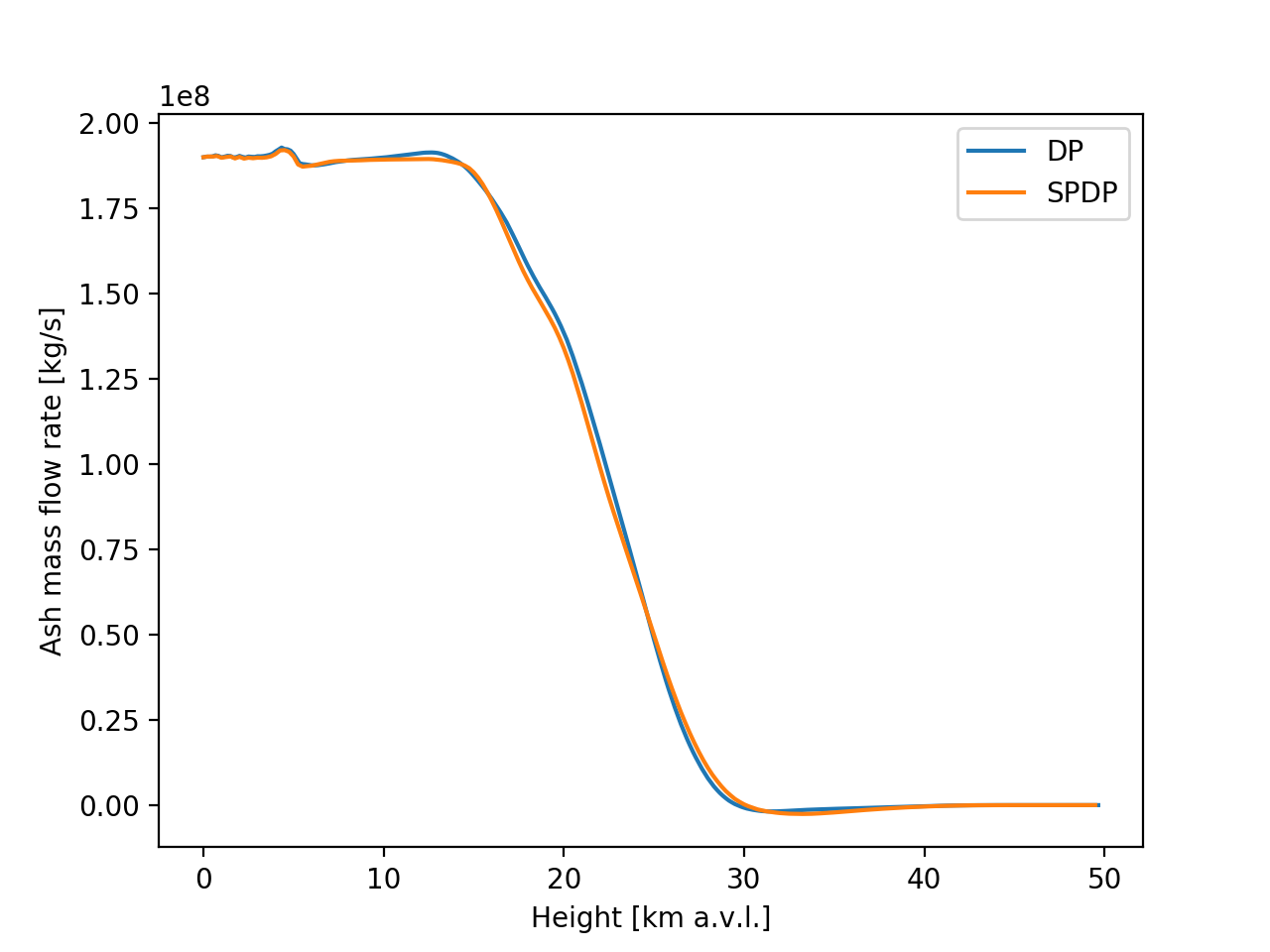}}
\caption{Numerical simulations of a volcanic plume with OpenFOAM-based solver \texttt{ASHEE} \citep{Cerminara2016}: a) snapshot of the simulation 9 minutes after the onset of the eruption; b) Comparison of vertically averaged profiles for the mass flow rate in double (DP) and mixed precision (SPDP).}
\label{fig:plume}
\end{figure}

In Fig.~\ref{fig:plume}a, we show a snapshot of the simulation 9 minutes after the onset of the eruption. The plume, described by an isosurface of ash concentration, reaches a maximum altitude of 30 km. The entire simulation covers 50 minutes of plume dynamics. In Fig.~\ref{fig:plume}b, we show vertical profiles of mass flow rate going upward inside the plume. They are obtained by averaging in time and horizontally in space over a 30 minutes time window. These profiles are used to compare the results in double and mixed precision. Mixed precision presents a good level of accuracy to reproduce the averaged properties of the simulated volcanic plume.

%% file: 2-performance_gain.tex
\section{Performance gain of reduced precision computations}
\label{sec:performance_gain}

Most CFD solver algorithms, as those implemented in OpenFOAM, can be divided into two main steps: the assembly of a matrix that results from the discretisation (Finite Volumes, in OpenFOAM) of the governing partial differential equations, and the solution of the linear algebra system that brings to the numerical solution of the equations. In some CFD libraries, all equations are merged in a single matrix and the resulting coupled system is solved at once. In OpenFOAM instead, each equation is solved separately (segregated strategy) and then coupled with the other ones using well-known iterative pressure- or density-based procedures (SIMPLE, PISO or PIMPLE; \citep{Ferziger2002}). 

In the following, we will refer to "matrix assembly" as the set of all the operations necessary for the construction of the matrices. The other main portion of the code, which we will refer to as "linear algebra", includes the set of operations aimed at solving the linear system generated during the "matrix assembly" phase. Finally, the last portion of the code that we will mention is the "flux correction" phase.
Passing from double to single precision in an ideal world means that all computational and communication costs involving floating point numbers are halved. Clearly, the parts of the solver involving integer numbers are not affected by the lower precision. When considering mixed precision, the computational gain is less obvious to be predicted. In this case, only the matrix assembly is performed in single precision whereas the linear algebra is still in double precision. The fraction of time spent in linear algebra may vary considerably for different applications (e.g. incompressible and compressible problems) as does the speedup. Moreover, the actual gain obtained with reduced precision depends on how much the application is memory bound and therefore on the details of the numerical setup (e.g. mesh size). 
Here we consider the lid-driven cavity and starting jet test cases to study and quantify the computational gain associated with reduced precision for incompressible and compressible CFD solvers (i.e. \texttt{icoFOAM} and \texttt{rhoPimpleFOAM}). The main reported metrics are the time spent for each part of the code in single and mixed precision, as compared to double precision.

A first analysis is conducted by changing the dimension of the computational domain and maintaining the number of cells per core constant (weak scaling analysis). This approach has been chosen to test the consistency of the observed gains while fixing the computational load and the memory bandwidth available per core. Subsequently, the test is repeated with an increased computational load to study its effect on the computational gain with reduced precision. The minimum amount of allocated resources considered is one socket (24 cores, on CINECA's Marconi machine (see \citep{Marconi} for hardware specifics) to keep a constant availability of DRAM bandwidth per core.

The results are reported in Tables \ref{tab:SPMPvsDPLidDriven} and \ref{tab:MPvsDPMiniApp}. The maximum gain provided by single precision compared with double precision is $2\times$.  For the lid-driven cavity in single precision, the computational gain is near to $2\times$  ($\approx1.9 \times$) for the test case with a larger computational load per processor (166k cells/core). Instead, for the lower computational load (41k cells/core), the gain obtained is smaller.
A deeper inspection reveals that this low value of the total gain is mainly related to the small gain of the linear algebra part, that in this case takes more than 90\% of the total time. Hence, as recently well explained by \citet{Zounon2022}, the 2$\times$ speedup for linear algebra is only obtained for larger problems. 
In fact, the gain of linear algebra is affected by the performance of the preconditioner, the Sparse Matrix Vector Multiplication (SpMVM) kernel and by the weight of the MPI communications and the slack time due to global synchronizations and not perfect load balancing (in single precision the MPI time is reduced but not halved and it is more than 30\% of the total time in the XL case).
Regarding the SpMVM kernel, since it is memory bandwidth-bound, the speed-up in theory is given by the ratio of the time required for matrix storage in double precision with respect to the lower precision. For a symmetric ordered LDU-COO matrix (native format in OpenFOAM) the speed-up (S) is given by the following formula:
\begin{align}
    S &= \frac{2*\text{nFaces}*\text{sizeOf}(\text{int32}) + (\text{nCells}+\text{nFaces})*\text{sizeOf}(\text{real64})}{2*\text{nFaces}*\text{sizeOf}(\text{int32}) + (\text{nCells}+\text{nFaces})*\text{sizeOf}(\text{real32})} \nonumber \\
    &= \frac{ 2 + \frac{ \text{nCells} }{ \text{nFaces} } }{ \frac{1}{2} ( 3 + \frac{ \text{nCells} }{ \text{nFaces}    })} \,,
\end{align}
where nCells and nFaces are the number of Cells and Faces respectively. For the Lid-driven cavity M the ratio is approximately 1.4. Our tests have shown that the solver gain is above this theoretical value except for the S case.\footnote{For the lid-driven cavity, we used the Conjugate Gradient preconditioned by a diagonal-based Incomplete Cholesky for symmetric equations and the Bi-Conjugate Gradient preconditioned by an incomplete LU factorization for asymmetric equations.} This result is in line with the recent work of \citet{Zounon2022} where they have studied the gain using single precision arithmetic in ILU and SpMVM kernels for a large set of sparse matrices. They found performances below $1.5\times$ in the solution phase of the application of the preconditioner and $1.5\times$ in the SpMVM kernel (using 10 cores). 

\begin{figure}[ht]
\centering
\subfloat[][\emph{Lid-driven cavity - Dimension M - 48 Cores}.]
{\includegraphics[width=1\columnwidth]{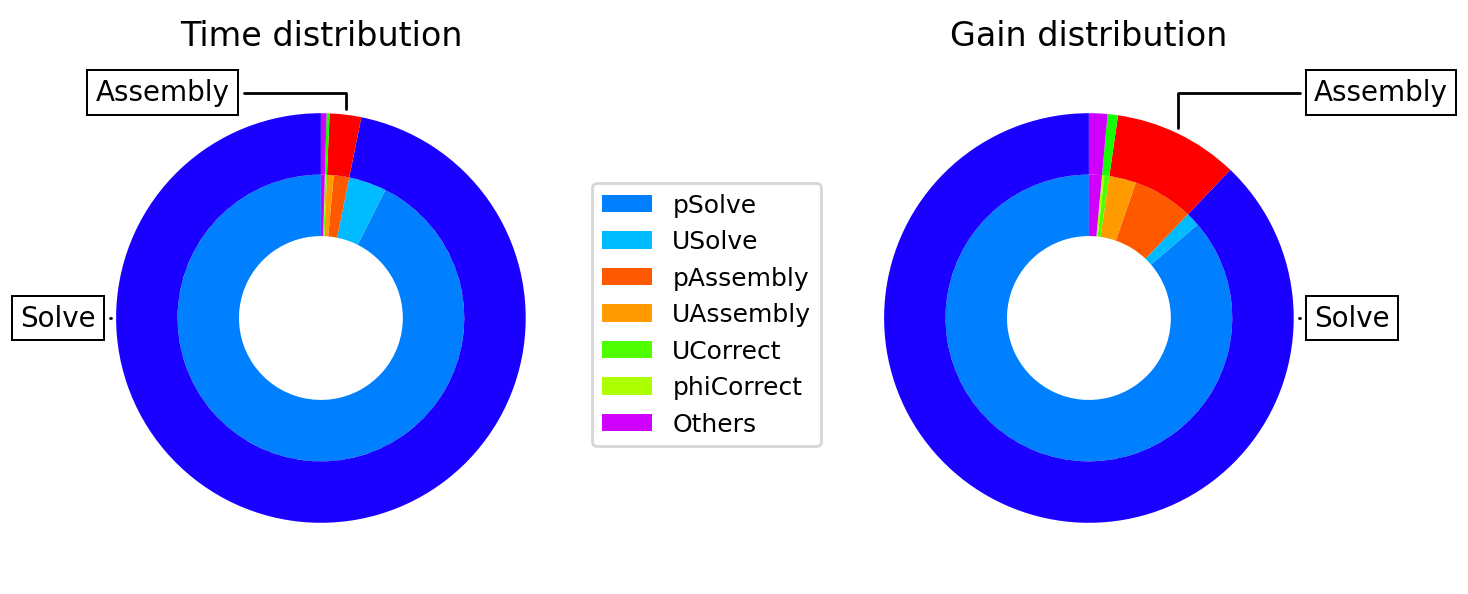}} \\
\subfloat[][\emph{Starting jet - Dimension M - 96 Cores}.]
{\includegraphics[width=1\columnwidth]{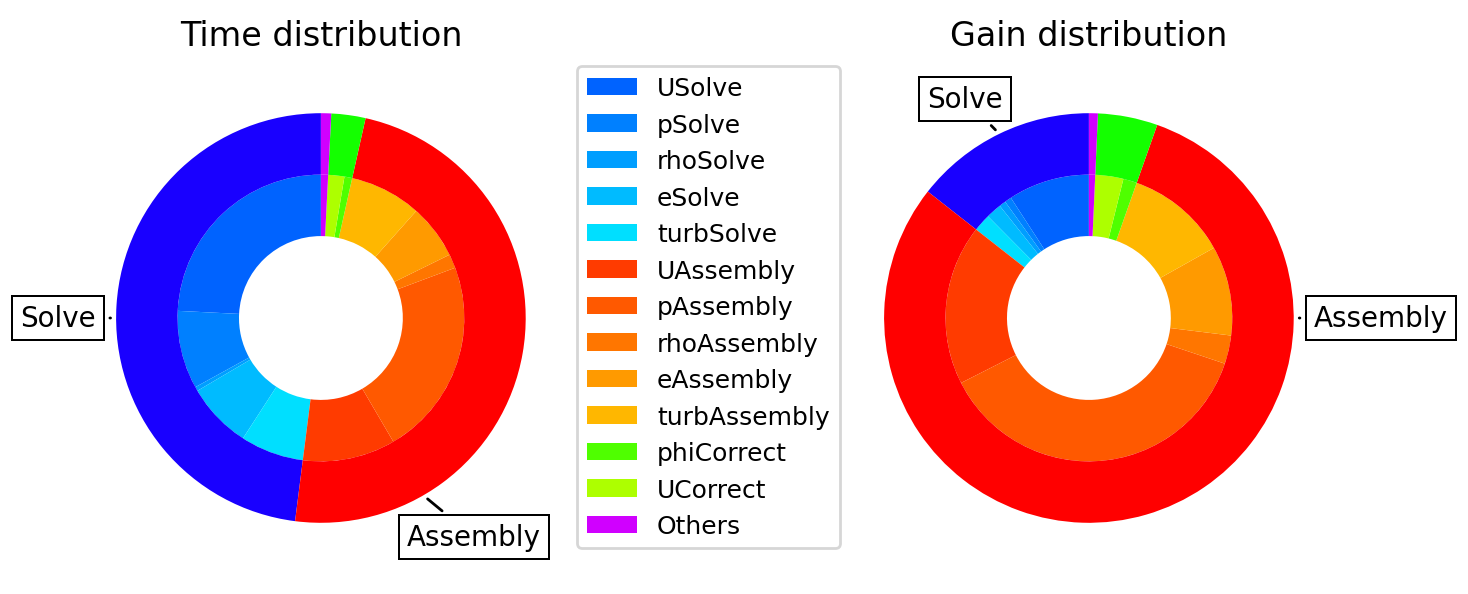}}
\caption{CPU-time distribution and performance gain using Mixed precision in the case M of the lid-driven cavity and the compressible starting jet.}
\label{fig:pieChart}
\end{figure}

As expected instead, for the mixed precision the gain obtained in matrix assembly and correction of the fluxes (single precision) is near two, while the gain in the resolution of the linear algebra (double precision) is near to 1.
Interestingly a small gain can also be observed in the solver phase for both the lid-driven cavity and the starting jet. 
Overall, the total gains (matrix assembly and solve) for the lid-driven cavity and starting jet are significantly different ($1.14\times$ versus $1.40\times$ respectively, in the M test-case with high computational load). The reason for this difference lies in the fact that the two applications have completely different time distribution between the matrix assembly and linear algebra phases (Figure \ref{fig:pieChart}). For the lid-driven cavity, most of the time (>90\%) is spent solving the linear algebra (mostly for the pressure equation), while for the starting jet the time is split roughly in half between the two phases. Hence, the gain obtained in the lid-driven cavity is mainly due to the small gain obtained in the solver phase. The total gain for the starting jet is mainly due to the halving of the time spent in the assembly phase.
Furthermore, from the Tables \ref{tab:SPMPvsDPLidDriven} and \ref{tab:MPvsDPMiniApp}, it can be noticed that the algebra solver phase percentage increases by increasing the number of core used. This could be attributed to the increased weight of the communication as the number of cores increases. In particular, this increase might be primarily ascribed to the global MPI calls contained in the solver phase (i.e. {\tt MPIAllReduce}). As expected, in mixed precision an increase of the time spent in linear algebra causes the total gain for both applications to decrease. 
\begin{table}[ht]
    \centering
    \includegraphics[width=1\columnwidth]{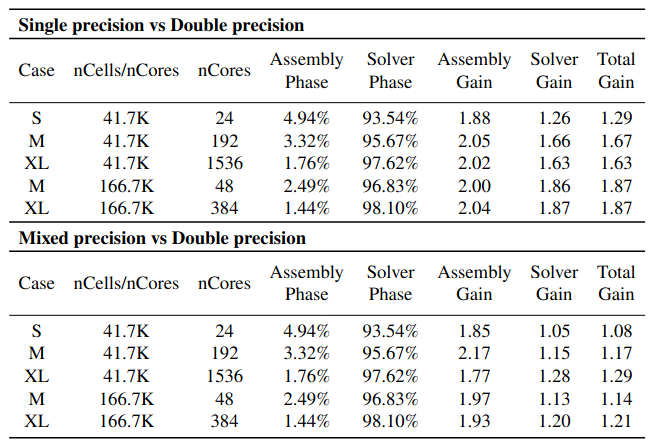}
    \caption{Lid-driven cavity: performance gain in weak scaling tests using Single and Mixed preci-
sion vs Double precision.}
    \label{tab:SPMPvsDPLidDriven}
\end{table}

\begin{table}[ht]
    \centering
    \includegraphics[width=1\columnwidth]{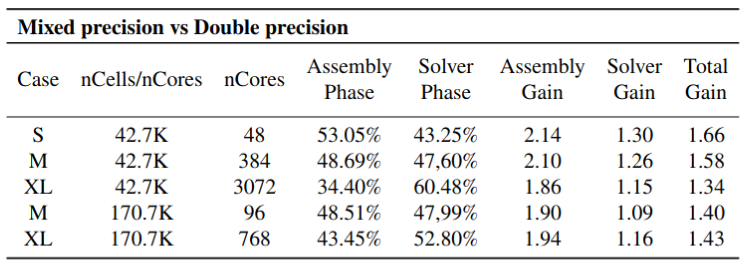}
   \caption{Starting jet: performance gain in weak scaling tests using Mixed precision vs Double precision.}
   \label{tab:MPvsDPMiniApp}
\end{table}

%% file: 3-scaling_analysis.tex
\section{Parallel computations and precision reduction}

\subsection{A theoretical model for strong scalability}

We here discuss a relatively simple theoretical framework that can be used to describe and analyse the scaling behaviour of an application on parallel machines, including the effect of reduced precision and hardware specifics (e.g. memory bandwidth or processor). The main outcome here is a model for the strong scalability, that is, the time gain (or the speedup) obtained by increasing the number of processors for a fixed problem size.

Let's start considering a generic numerical problem that is solved using a variable number of processors $P$ (MPI processes).
On a first approximation, the total time ($T$) required by a single processor ($p$) to solve its own sub-problem is a function of: the time to transfer the data from the RAM memory of a process to another (intra-node and inter-node), the time to transfer the data from the RAM memory of a process to the registries, the time to process the data. Assuming that these three events occur in three non-overlapped time phases (e.g. \citet{Succi2019}), the total time is then given by the sum of three terms:
\begin{equation}
    T = \frac{F}{\dot{F}} + \frac{B}{\dot{B}} + \frac{C}{\dot{C}} \label{eq:totalTime}
\end{equation}
where $F$ is the sum of floating point instructions done by the processor in FLOP (both in single and double precision), $\dot{F}$ the processor speed in FLOP/s, $B$ the amount of data, in Bytes, to be transferred from the RAM to the registries, $\dot{B}$ the memory bandwidth in Byte/s, $C$ the amount of data, in Byte, to be transferred from the RAM memory of a process to another, and $\dot{C}$ the network bandwidth, in Byte/s. Now, in order to understand the strong scaling behaviour of an application we need to understand how each term in eq. \ref{eq:totalTime} scales with the number of processors. 

The first term ($F/\dot{F}$) represents the time required for pure calculations (no communication). Increasing the total number of cores simply reduces in general the computational work ($F$) assigned to the single processor. However, determining its exact behaviour is not an easy task. The actual load balancing and the ratio between the number of internal and boundary cells is a function of the algorithm used for domain decomposition. Moreover, the number of iterations for the preconditioned linear solver to converge can increase when switching from a serial to a parallel implementation. Therefore here we simply consider that a perfectly robust solver is used (i.e. the number of iterations does not change from the serial to the parallel case), and the number of internal cells is much higher than boundary cells and is the same among the processors (perfect load balancing). In this case, $F$ can be modelled simply as inversely proportional to the total number of MPI processors ($P$): 
\begin{equation}
 \label{eq:Fdef}
    F = \frac{k_F}{P}
\end{equation}
where $k_F$ is a constant expressing the total number of FLOP, in single and double precision, required to solve the entire numerical problem.
By definition, the processor speed ($\dot{F}$) is independent of the number of processors and is expressed as the product of the average CPU frequency ($f_{CPU}$ in Hz) and the Floating Point Instructions retired Per Cycle ($k_{IPC}$)
\begin{equation}
         \label{eq:Fdotdef}
    \dot{F} = k_{IPC} f_{CPU} = k_{\dot{F}} 
\end{equation}
 where $k_{IPC}$ and  $f_{CPU}$, hence also $k_{\dot{F}}$, are assumed to be constant.

For the same reasoning described for $F$, on a first approximation also $B$ can be modelled as inversely proportional to the number of processors:
\begin{align}
    B = \frac{k_B}{P} \label{eq:Bdef}
\end{align}
where the constant $k_B$ represents the total amount of data for communications with memory for all processors.
Regarding the memory bandwidth ($\dot{B}$), we need instead to distinguish the intra-node from the inter-node case.

Within a computing node, the RAM memory channels are shared among the cores and memory bandwidth contention cannot be avoided. In such a case, the memory bandwidth increases with the number of cores until saturation (Appendix Figure A1a). The behaviour is almost linear for a few cores and then starts to bend towards the asymptotic value. The memory bandwidth per core slowly decreases until saturation and then drops (Appendix Figure A1b). We modelled this behaviour with an expression for the inverse of the memory bandwidth equivalent to a first-order Taylor polynomial function of the number of processors (eq. \ref{eq:Bdotdef_intra}).

When considering multi-node simulations, the RAM channels are still shared inside the node but not among the nodes. Therefore, in an inter-node strong scaling, we can assume that the memory bandwidth for each processor is constant and equal to the value for the single node (considering the full node as a reference and multiples of a single node to measure the speedup).
However, the presence of cache levels (L1, L2 and L3) increases the value of $\dot{B}$ as the number of processors grows, due to an increase in the cache efficiency due to an increase in the total cache size available in the system \citep{Benzi09}. 
The bigger the number of processors, the smaller the size of the data representing our discrete problem, until it fits into the cache. In this limit case, we get the maximum memory bandwidth of the system.
We modelled this performance augmentation as a linear function (Eq. \ref{eq:Bdotdef_inter}). As a result, for the intra-node and inter-node scaling test we have two different equations expressing the memory bandwidth available for each processor:
\begin{align}
    \frac{1}{\dot{B}_{intra}} &= \frac{1}{k_{\dot{B}}(P)} = k_1 + k_2 (P - 1) \label{eq:Bdotdef_intra} \\
    \dot{B}_{inter} &= k_{\dot{B}}(P) = k_3 (1 + k_4 \cdot (\frac{P}{n} - 1)) \label{eq:Bdotdef_inter}
\end{align}
where $n$ is the number of cores in a node, $k_{\dot{B}}$ is the memory bandwidth and $k_1$, $k_2$, $k_3$ and $k_4$ are model constants. When $P=1$ in the intra-node case and $P=n$ in the inter-node case, $k_1$ and $k_3$ coincides with the memory bandwidth of the serial case.

Finally, we need to consider the third term  of Eq. \ref{eq:totalTime}, the time spent by the single process in sending and receiving data from other processes. As it is common in parallel CFD applications,  OpenFOAM make use of MPI (e.g. Open MPI) for both point-to-point and collective communications. Asynchronous Point-to-Point communications are required for instance in the assembly phase or in a matrix-vector multiplication, while collective communications are used in {\tt MPIAllreduce} functions required by linear algebra solvers.  
The total time spent in MPI communications depends on the number of cores in a strong scaling experiment. Typically, as the number of cores is sufficiently small, the wall-time fraction spent in the MPI library  ($\frac{C}{\dot{C}}$ in Eq. \ref{eq:totalTime}) is very small, almost negligible. Instead, MPI overhead starts to be significant for a large number of cores or, to be more precise and general, when the number of cells per core is small.
For instance, in our simulations with OpenFOAM the MPI communications take less than the 5\% of the total time  if the number of cells per core is above around 150k. However, when the number of cells per core approaches around 20k, MPI communications become more and more important and may take up to 50\% of the total time for many cores (Figure A3).
However, even if the relative weight of MPI increases as the number of core increases, the absolute total time spent in MPI decreases.
Indeed, in our test when the number of cells per core is sufficiently small the time spent in MPI communications becomes important and approaches an asymptotic value (Figure A3). As a result, we model the third term of Eq. \ref{eq:totalTime} as:
\begin{equation}
    \frac{C}{\dot{C}} \asymp k_C
    \label{eq:MPItime}
\end{equation}
where $k_C$ is a constant. Let us enforce that this approximation cannot be used in a strong scaling for runs with a large number of cells per core. However, in this case the time spent in MPI communications is not significant with respect to the total time and as a first approximation can be simply neglected.
A more accurate approach would require to model the exact behaviour of the dominant type of MPI calls (e.g. {\tt MPIAllReduce}) as a function of the number of cores/cells per core as described in Appendix A. Here for practical reasons we simply  express  $\displaystyle \frac{C}{\dot{C}} \approx k_C w(P)$ where $w(P)$ is a weight function that varies from zero to one passing from runs with large to small number of cells per core.

Now, using eq. \eqref{eq:Fdef}, \eqref{eq:Fdotdef} and \eqref{eq:Bdef} one can write the equation for the time required to solve the problem by a single processor as a function of the number of processors ($P$):
\begin{equation}
    T_{intra}(P) = \frac{k_F}{P k_{\dot{F}}} +  \frac{k_B}{P k_{\dot{B}}(P)} + w(P)k_C
    \label{eq:T-general}
\end{equation}
where  $k_{\dot{B}}(P) = 1/(k_1 + k_2 (P - 1))$ for the intra-node and
$k_{\dot{B}}(P) = k_3(1 + k_4 (nP - n))$ for the internode.   
It is then straightforward,  obtain the formula for the  speedup, ($S(P) = T(1)/T(P)$) that is commonly measured in strong scaling  experiments for both the intra and inter-node test cases: 
 \begin{equation}
    S(P) = \frac{k_F /  k_{\dot{F}} + k_B / k_{\dot{B}}(1) + w(1)k_C}{ k_F/ ( P \cdot k_{\dot{F}} (P) ) + k_B/(P \cdot k_{\dot{B}}(P)) + w(P)k_C} 
    \label{eq:S-general}
\end{equation}
   The above equation is relatively general and, given that all the parameters are known, could be used to predict the scaling behavior of an application, including the effect due to a change in floating point precision and hardware specifics (e.g. memory bandwidth). In fact,  a precision reduction would change the value of most of the parameters appearing in the equation. However, a precise determination of all these constants is beyond the scope of the present work. Instead, in the next sections we will focus on showing how this model can be used to provide qualitative description  of  the speedup observed in strong scaling experiments of real CFD applications, such as those implemented in OpenFOAM.

\subsection{Intra-node scalability}

\begin{figure}[ht]
 \centering
\subfloat[][\emph{}]
{\includegraphics[width=0.5\columnwidth]{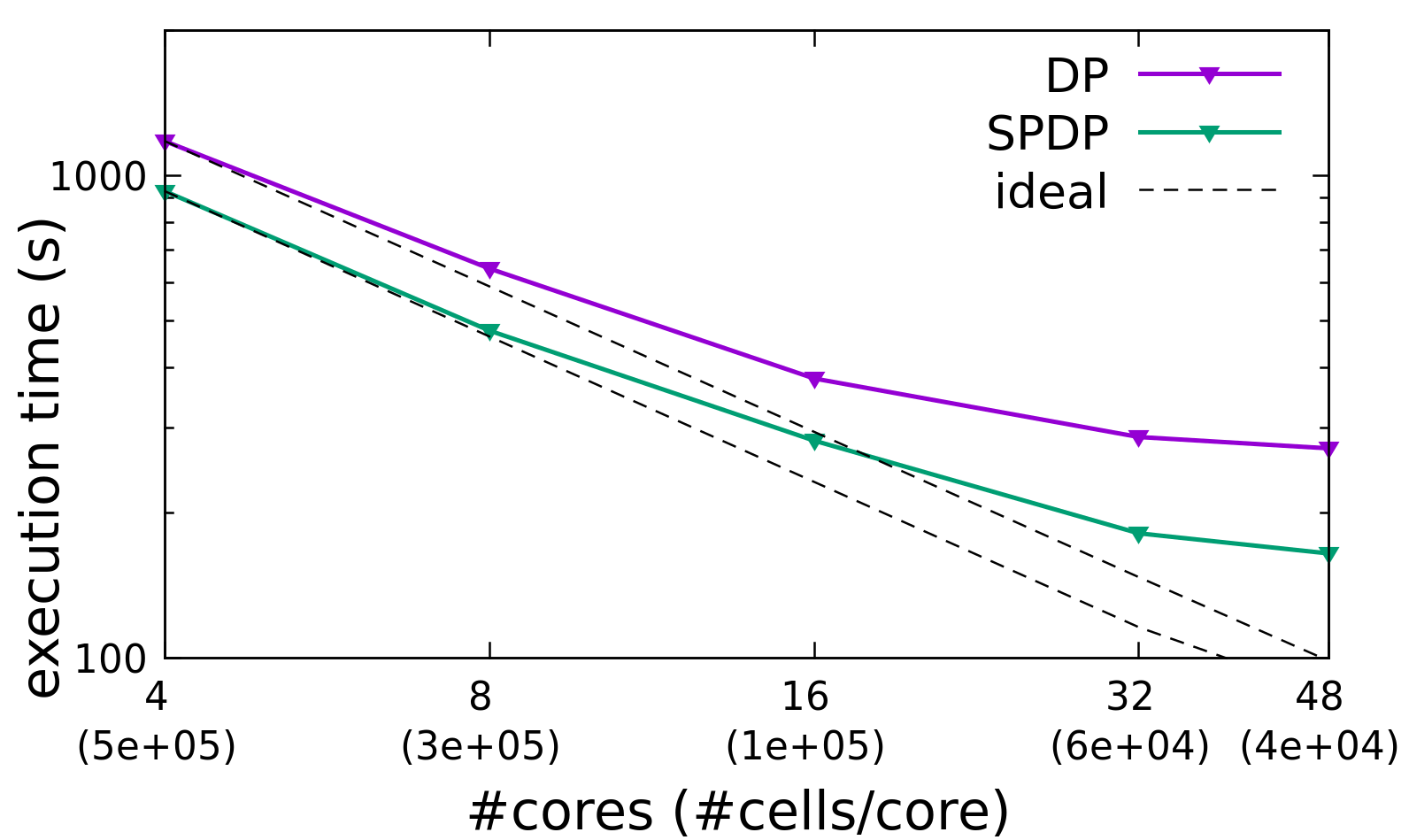}} 
\subfloat[][\emph{}]
{\includegraphics[width=0.5\columnwidth]{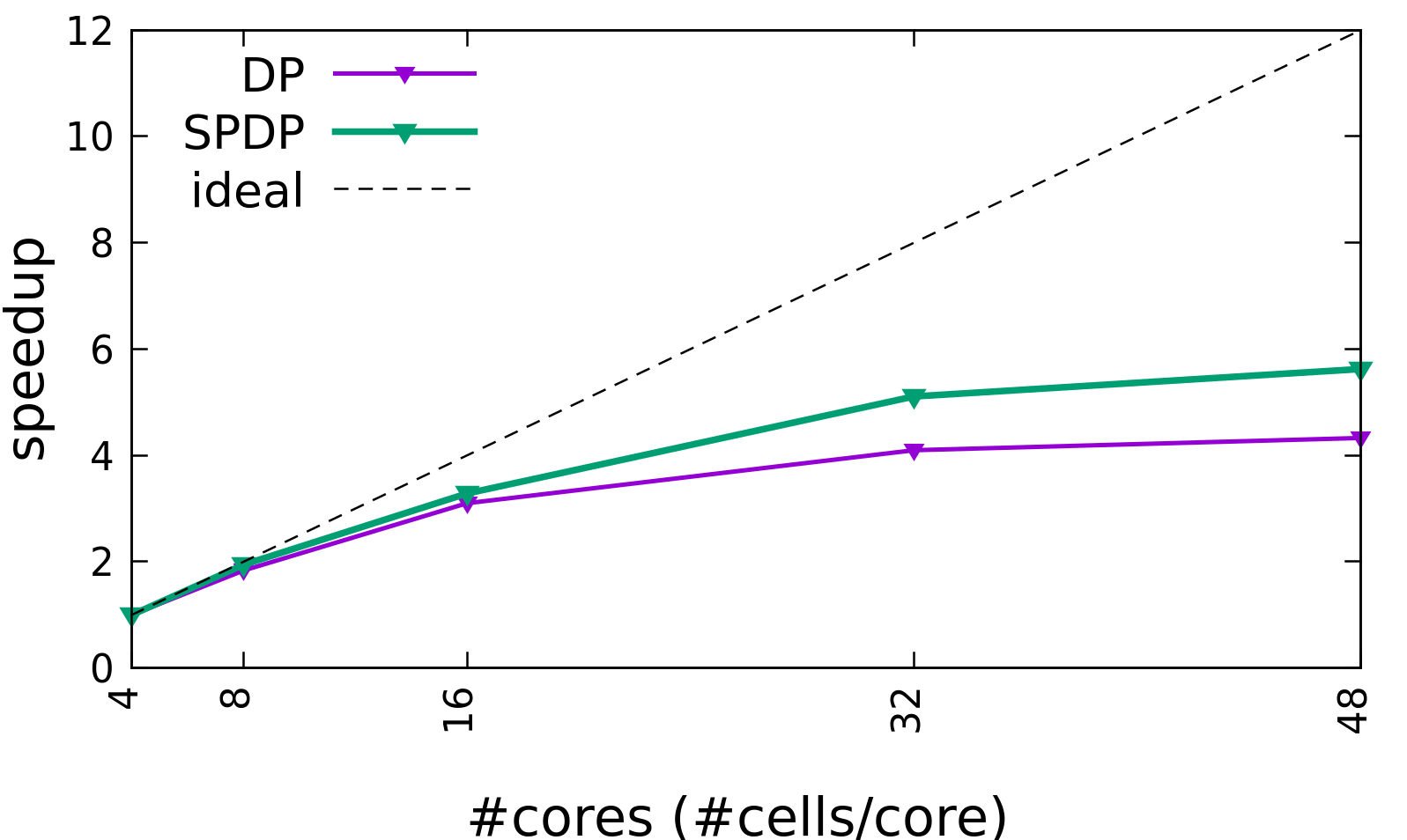}}
 \\
\subfloat[][\emph{}]
{\includegraphics[width=0.5\columnwidth]{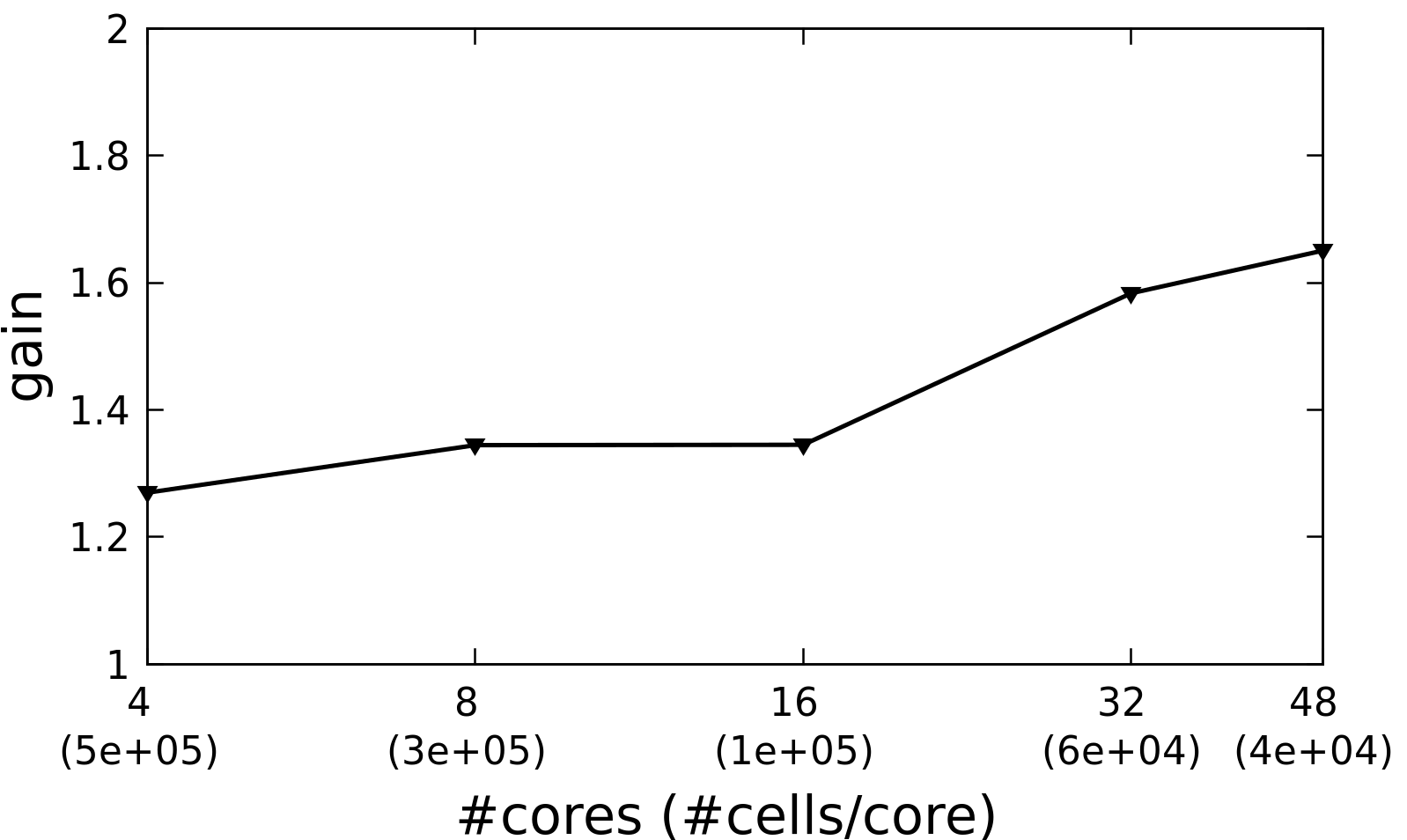}}
\subfloat[][\emph{}]
{\includegraphics[width=0.5\columnwidth]{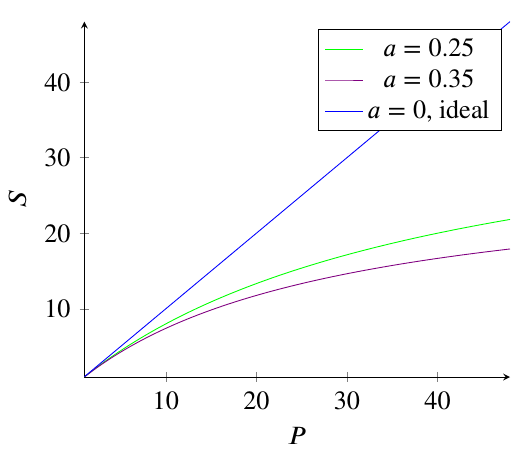}} \\
\caption{Intra-node strong scaling for the starting jet using a mesh size of $\approx 2\times10^6$ cells performed on Marconi machine (a,b,c). The speedup obtained with the theoretical model (d), is also reported for a qualitatively comparison with the experimental one (b).}
\label{fig:intranode-jet}
\end{figure}

We now demonstrate and analyse the beneficial effects of using reduced precision on the scaling properties at the intra-node level. A scaling experiment with the starting jet test case compiled in double and mixed precision is considered. The size of the test case is relatively large to keep the number of cells per core above the threshold of 40k. In this way, memory communications are expected to play a significant role whereas the contribution from MPI communications should be negligible. 
The observed speedup deviates quite quickly from the ideal behaviour for both double and mixed precision (Figure \ref{fig:intranode-jet}), clearly indicating an inefficient use of the computational resources. However, reduced precision helps to improve both the scalability and wall time of the application. In particular, the mixed precision scales better than double precision and provides a computational gain (up to $1.6\times$) that increases across the node.  

To understand the reasons behind this behaviour, we now reconsider the scaling model described in the previous paragraph (eq. \ref{eq:S-general}).
Neglecting the effect of MPI communications ($w(P)=0$) and applying the approximation for $k_{\dot{B}}$ described in Eq. \ref{eq:Bdotdef_intra},
after some algebraic manipulation we obtain a simplified form of eq. \ref{eq:S-general}:
\begin{equation}
\label{eq:Sintranode}
    S(P) = \frac{1}{ \frac{1}{P} \cdot (1 + a \cdot (P - 1))}
\end{equation}
where the constant $a$ is defined as
\begin{equation}
\label{eq:aDef}
 a = (k_B \cdot k_2) / (k_F/k_{\dot{F}}(1) + k_B/k_{\dot{B}}(1))
\end{equation}
Eq. \eqref{eq:Sintranode} is plotted in Figure \ref{fig:intranode-jet} for different values of $a$. It is clear that the lower the value of $a$ the closer the curve is to the ideal speed-up. The overall effect of decreasing the parameter $a$ therefore is similar to the effect of the precision reduction observed in our scaling tests (Figure \ref{fig:intranode-jet}). In other words, one may argue that mixed precision scales better than double precision at the intra-node level thanks to a lower value of the parameter $a$. The decreased value of this parameter with reduced precision can be easily justified by the reduction of the amount of data to be transferred during matrix assembly (lower value of $k_B$) and the increase of the transfer rate of data due to a reduction of memory stalls (lower value of $k_2$), as also confirmed by the level 1 of the TMAM analysis (Appendix Figure A2). We instead assumed that the other parameters of Eq. \ref{eq:aDef} are not influenced by the type of precision: the number of instructions is the same, as a consequence of the same algorithm;  the memory bandwidth and the CPU frequency are identical due to the same hardware; IPC is also the same since we assume that floating-point operations are implemented natively in the hardware for both single and double precisions (and vectorization is not present). Thus, the decrease of scalability observed at the intra-node level seems related to memory communication and hence it can be improved by using reduced precision. 


\subsection{Inter-node scalability analysis}

Similarly to the previous section, we discuss the effect of precision reduction on the inter-node scalability by considering the starting jet case compiled in double and mixed precision (OpenFOAM version v2006). In the present scaling analysis, we consider the execution time of the application without pre-processing (e.g. mesh generation) and initialisation time delay at the beginning of the simulation. \footnote{The initialisation time is usually very small compared with the total execution time, especially for long production runs.  However, for our setup, the execution time is also quite small and therefore it might become comparable with the initialisation time.}  

\begin{figure}[ht]
 \centering
\subfloat[][\emph{}]
{\includegraphics[width=0.5\columnwidth]{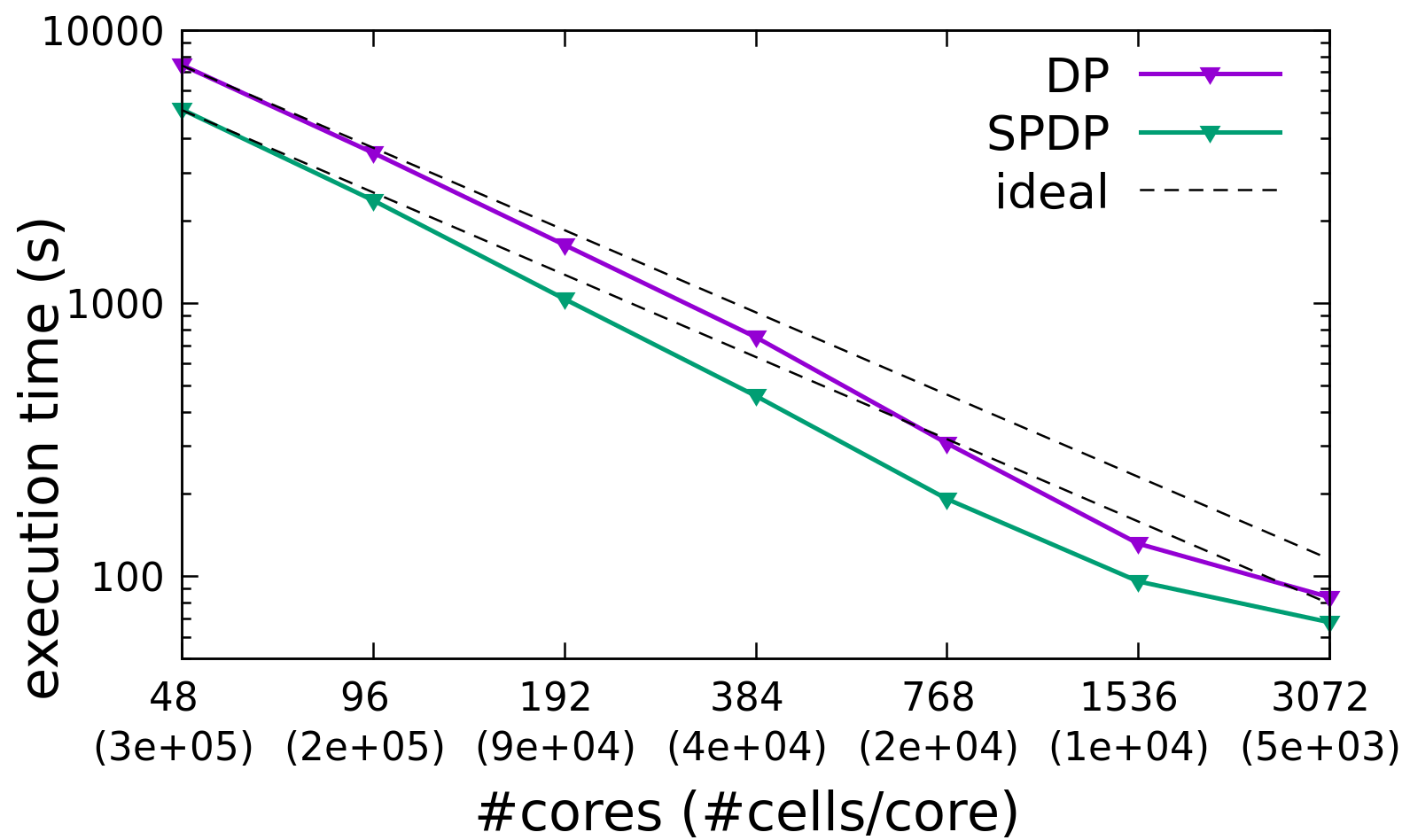}} 
\subfloat[][\emph{}]
{\includegraphics[width=0.5\columnwidth]{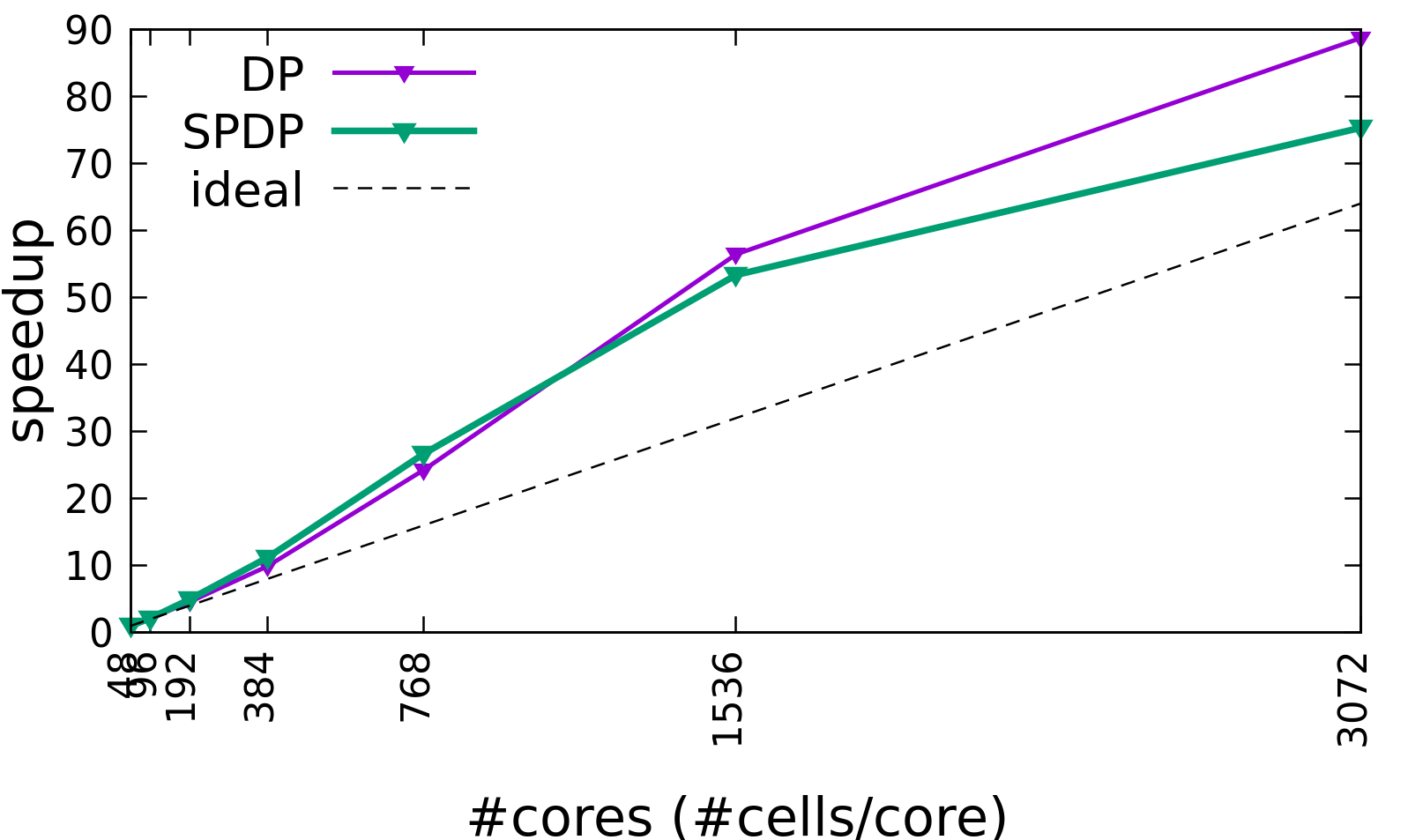}}
 \\
\subfloat[][\emph{}]
{\includegraphics[width=0.5\columnwidth]{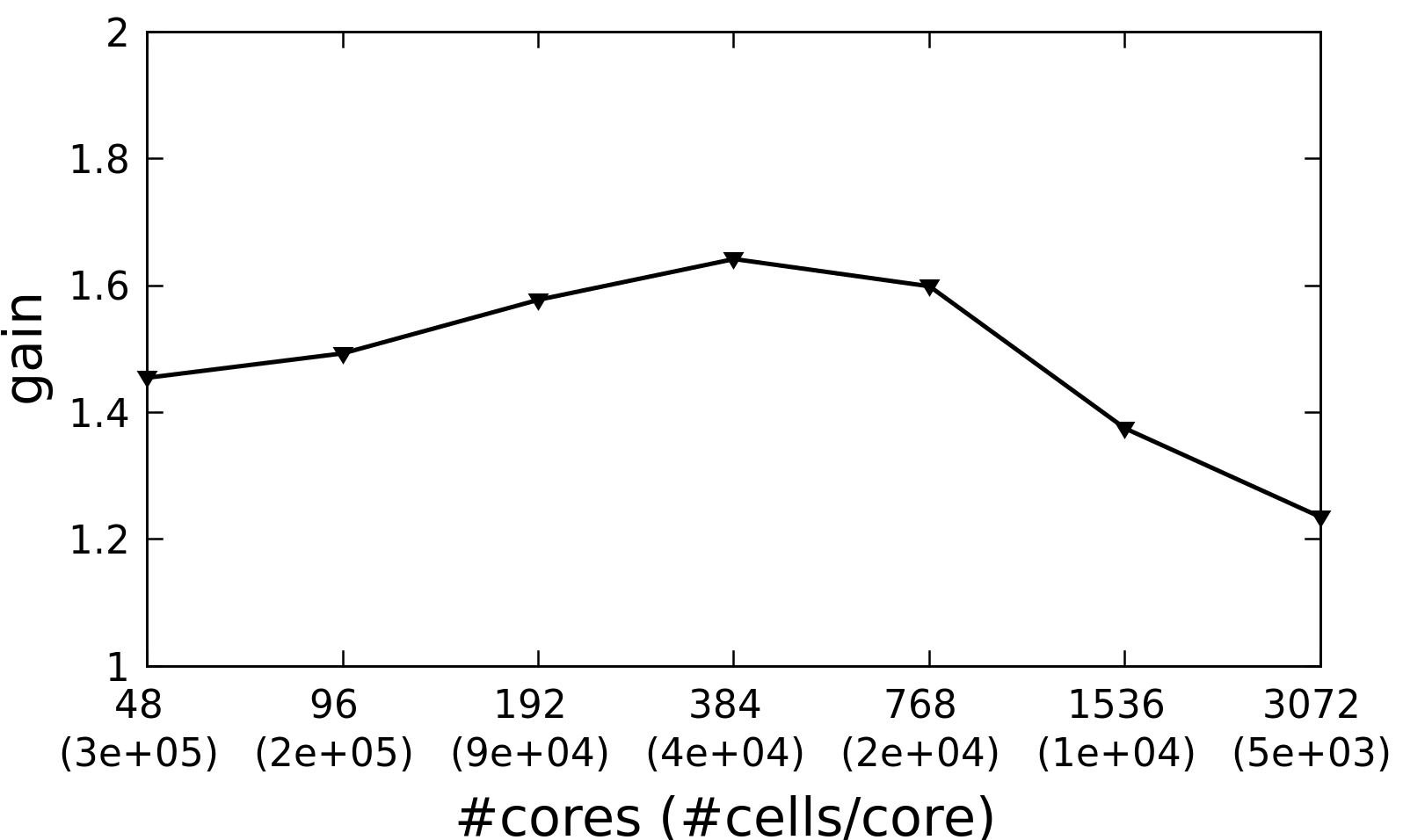}}
\subfloat[][\emph{}]
{\includegraphics[width=0.5\columnwidth]{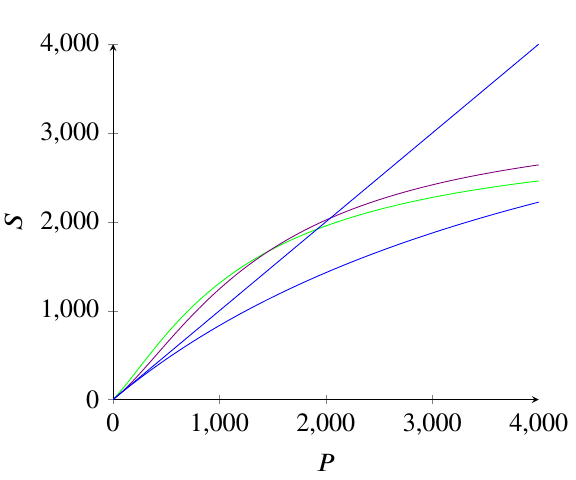}} \\
\caption{Inter-node strong scaling for the starting jet using a mesh size of $\approx 16\times10^6$ cells performed on Marconi  (a,b,c). The speedup obtained with the theoretical model (d), is also reported for a qualitatively comparison with the experimental one (b).}
\label{fig:internode-jet}
\end{figure}

The results are reported in (Figure \ref{fig:internode-jet}). Both full and reduced precision runs display very similar speedup, with a clear superlinear behaviour up to $10^4$ cells/core (Figure \ref{fig:internode-jet}a,b). Below this threshold, MPI communications start to be more important, the scalability quickly deteriorates and double precision scales better than mixed precision.  Nonetheless, it is important to notice that mixed precision is always faster than double precision. The computational gain in fact steadily increases from $1.4\times$ up to $1.6\times$ at $20^4$ cells/core and then rapidly decreases to $1.2\times$, in the MPI-bounded region (Figure \ref{fig:internode-jet}c).  Hence, as expected, mixed precision does not directly improve the parallel efficiency of the application as in the intra-node case, but it is still computationally convenient up to the number of cells/core where the application present a good parallel efficiency.

For the inter-node case, the MPI overhead cannot be neglected and has to be considered. For simplicity here, in the third term of Eq. \ref{eq:T-general}, we set $w(P)=1$ for multiple node runs and $w(1)=0$ for the reference case (single node). As a result, Eq. \ref{eq:S-general} can be written as  
\begin{equation}
    S(P) = \frac{k_F /  k_{\dot{F}}(1) + k_B / k_{\dot{B}}(1)}{ k_F/ ( P \cdot k_{\dot{F}} (P) ) + k_B/(P \cdot k_{\dot{B}}(P)) + k_C} \,.
\end{equation}
Then, assuming $k_{\dot{F}}$ to be constant and using equation \eqref{eq:Bdotdef_inter}, after some algebraic manipulation we get 

\begin{equation}
\label{eq:Sinternode}
    S(P) = \frac{1}{\frac{b_1}{P} + \frac{b_2}{P \cdot (1 + k_4 \cdot (\frac{P}{n} -  1))} + b_3}
\end{equation}
where the following definitions hold
\begin{align}
    b_1 &= \frac{1}{1 + \frac{k_B \cdot  k_{\dot{F}}}{k_{\dot{B}}(1) \cdot k_F}  } \,, \\
    b_2 &= \frac{1}{1 + \frac{k_F \cdot  k_{\dot{B}}(1)}{k_{\dot{F}} \cdot k_B}  } \,, \\
    b_3 &= \frac{k_C}{k_F /  k_{\dot{F}(1)} + k_B / k_{\dot{B}}(1)} \,.
\end{align}
Eq. \eqref{eq:Sinternode} is plotted in Figure \ref{fig:internode-jet}d. A good agreement between the model and the curve plotted in Figure \ref{fig:internode-jet}b) has been found for the Marconi cluster using the parameters $b_1=0.3$, $b_2=0.5$, $k_4=0.003$, $b_3=0.00032$ for the mixed precision and $b_1=0.22$, $b_2=0.85$, $k_4=0.002$, $b_3=0.00025$ for the double precision.
For the same reasons discussed for the intra-node case, the constant $b_3$ is higher in mixed precision with respect to double precision. Regarding the variation of the parameters $b_1$ and $b_2$, it depends on the ratio $\frac{k_B \cdot  k_{\dot{F}}}{k_{\dot{B}} \cdot k_F}$. If the ratio $\frac{k_{B}}{k_{\dot{B}}}$ decreases more than $\frac{k_{F}}{k_{\dot{F}}}$, the variation of $b_2$ is negative and of $b_1$ is positive, and viceversa.
Since we can reasonably assume that the variation of $\frac{k_{F}}{k_{\dot{F}}}$ is less pronounced than $\frac{k_{B}}{k_{\dot{B}}}$ (in the previous paragraph we assumed no variation for the term $\frac{k_{F}}{k_{\dot{F}}}$), $b_1$ and $b_2$ are higher and lower in the mixed precision case, respectively.
Regarding the parameter $k_4$, we can expect a higher value of this parameter in mixed precision due to a stronger cache effect when working with floats.

In conclusion, the two phenomena that influence the scaling plot are super-linearity and MPI communication. The super-linearity is due to the presence of levels of cache that, as the number of nodes grows, increasingly reduces the memory bandwidth bottleneck and hence provides an additional speedup. The super-linearity is slightly more pronounced in mixed precision at a low number of cores. When the MPI communication starts to play a significant role, the mixed precision scales worst. The reason is that the MPI time is not affected by the mixed precision since most MPI calls are from the linear algebra solvers which is double precision. Moreover, since the MPI time is related to synchronization lags instead of the amount of data to be exchanged by processors, the same conclusion may be valid also if single precision is used everywhere in the code. 

%% file: 4-GPU.tex
\section{The role of mixed precision in a hybrid CPU-GPU implementation}

In the last decade, a number of GPU implementations of OpenFOAM have been attempted by individual research groups with different strategies and different results in terms of computational performance (e.g. \cite{Malecha2011,Alonazi2015,Krasnopolsky2016}). However, none of these attempts have been included yet in the official OpenFOAM distributions.

More recently, \citet{Bna2020} and \citet{zampini2020} have developed an interface library named PETSc4FOAM that has been included in the OpenFOAM official release as a module. PETSc4FOAM extends the list of available linear solvers with the ones embedded in PETSc and other packages(e.g. HYPRE and ML), with and without GPU support. 
It basically works with any OpenFOAM solver and allows the user to take advantage of GPU hardware without requiring any change to the source code.
Certainly, only the solution of the linear system can be offloaded to the GPU, the assembly of the matrix is still performed on CPU using the original OpenFOAM routines. Therefore this hybrid CPU-GPU strategy provide a speedup that primarily depends on the fraction of time spent in linear algebra by a specific application.
For instance, roughly 90\% of the computational work ("solve" in figure \ref{fig:pieChart}a) can be accelerated on GPUs for the Lid-driven cavity solved with \texttt{icoFoam}. This fraction reduces to 50\% (figure \ref{fig:pieChart}b) when considering the starting jet test case with \texttt{rhoPimpleFoam}. Morevoer, one has also to consider the time spent in CPU-GPU communication that may have a significant cost. OpenFOAM applications use a segregated solution strategy, that means that each equation is solved separately and then coupled through iterative procedures. The time spent in periodic transfer of the LDU matrix values to the GPU, every time a linear system has to be solved, may therefore increase dramatically with the number of equations to be solved.
Nevertheless,  the use of mixed precision may come in to rescue this hybrid CPU-GPU strategy.
Indeed, with the mixed precision, the amount of data involved in CPU-GPU communication is halved and matrix assembly on CPU becomes twice as fast. Moreover, if the matrix sparsity pattern does not change from one iteration to the next one, it can be cached on GPU and only the matrix values have to be offloaded (in single precision, in the SPDP case). 

If we neglect the time spent for the data transfer and the conversion\footnote{The matrix has to be converted from the native LDU-COO format to CSR each time the solver is called. This operation is performed efficiently on GPU using the radix sort algorithm.}, we can approximately write the total time as the sum of the time spent in the solution of the linear system and the time spent in the assembly of the linear operators. The default in OpenFOAM is to use double precision for all the computations and a solver based on CPU. In the formula we have
\begin{equation}
    T_{tot} = T_{solver\_{CPU}} + T_{assembly\_{DP}}    
\end{equation}
The fraction of time spent in the assembly is defined as 
\begin{equation}
    F = \frac { T_{assembly\_{DP}} } { T_{tot}  }
\end{equation}
As we already said, the hybrid CPU-GPU approach has no impact on the assembly loop in terms of computational time but it can be easily combined with mixed precision. On the contrary, mixed precision has no impact on the solution of the linear system but strongly influences the execution time of the matrix assembly part. 
Therefore, by defining the speedup obtained with the use of GPU linear algebra solvers and mixed precision  as: 
\begin{equation}
    S_{GPU} = \frac{ T_{solver\_{CPU}} }{ T_{solver\_{GPU}} } \qquad S_{P} = \frac{ T_{assembly\_{DP}} }{ T_{assembly\_{SPDP}} }
\end{equation}
 one can apply Amdahl's law and predict the total speedup of mixed precision and GPU linear algebra as:
 
\begin{equation}
\label{eq:formula_gain}
    S = \frac{T_{solver\_CPU} + T_{assembly\_DP} }{T_{solver\_GPU} + T_{assembly\_SPDP}} = \frac{1}{ \frac{1}{S_{GPU}}(1 - F) + \frac{1}{S_{P}} F}
\end{equation}

In Table \ref{tab:SPDP+GPUvsDP} we report the total Execution Time and the total Gain of the starting jet use-case using a different combination of precision and number of equations offloaded to the GPUs.  In particular, here we use a recent extension of PETSc4FOAM \citep{Martineau2021} that provides the possibility to use NVIDIA algebra solvers with GPU support of the AMGx library \citep{AmgX2015}. The best gain is achieved using the mixed precision and the solution of all equations on GPU. The gain can also be estimated using formula \eqref{eq:formula_gain}. In section \ref{sec:performance_gain} we showed that a reasonable value for $F$ and $S_{P}$ is $0.5$ and $2$, respectively. Using these values, the estimate of the total gain achieved using only the mixed precision is $1.333$ (the same value in Table \ref{eq:formula_gain} is 1.31). If the linear equations are solved on GPU, a reasonable value for $S_{GPU}$ taken from our experiments is $3$, which gives us an estimate of the total gain equal to 2.4 (compared to 2.43 in Table \ref{eq:formula_gain}).

\begin{table}[h]
\begin{tabular}{cccc}
\toprule
\multicolumn{4}{l}{\bf{Mixed precision vs Double precision for CPU-GPU OpenFOAM}} \\
\toprule
\thead{Solver}  & \thead{Precision} & \thead{Execution Time} & \thead{Gain} \\
\midrule
PBiCG.DILU  & DP & 1744 & - \\
PBiCG.DILU  & SPDP & 1329 & $1.31\times$ \\
\thead{PBiCGStab.JACOBI\_L1 \\ Equation (p) on GPU } & SPDP & 1237 & $1.41\times$ \\
\thead{PBiCGStab.JACOBI\_L1 \\ Equation (p|U) on GPU } & SPDP & 931 & $1.87\times$ \\
\thead{PBiCGStab.JACOBI\_L1 \\ Equation (p|U|e) on GPU } & SPDP & 833 & $2.09\times$ \\
\thead{PBiCGStab.JACOBI\_L1 \\ All equations on GPU } & SPDP & 733 & $2.38\times$ \\
\thead{PBiCGStab.BLOCK\_JACOBI \\ All equations on GPU } & SPDP & 718 & $2.43\times$ \\
\bottomrule
\end{tabular}
\caption{\label{tab:SPDP+GPUvsDP} Execution time and gain of the starting jet benchmark using different types of algebra solver and a single node of Marconi100 \citep{Marconi100}}
\end{table}

%% file: 5-conclusions.tex
\section{Discussions and conclusions}

We have analysed the effect of reduced precision on different test cases considering incompressible, compressible as well as multiphase flow problems.
Overall, single precision provides numerical solutions that are in good agreement with double precision or the theoretical ones 
for test cases where the flow is laminar (the lid driven cavity and the shock tube benchmark). Single precision therefore may be expected to be sufficiently accurate for most laminar fluid problems except for those cases where the initial conditions require double precision or the solution is required with a particular degree of accuracy.
When considering turbulent flows, single precision seems not to be an option instead. In this case, the nonlinear terms in the governing equations amplify the numerical noise, compromising the convergence of linear algebra solvers.
In particular, \texttt{rhoPimpleFoam} and \texttt{ASHEE} solvers have not converged to physical solutions when considering the dynamics of decaying isotropic turbulence,  starting compressible jet and turbulent multiphase flow of a volcanic plume. However, for these cases, one may rely on the implementation of a mixed precision concept in which linear algebra solvers work in double precision. In this case, all the aforementioned computations converge to consistent physical solutions. The spectrum of the kinetic energy of decaying isotropic turbulence in mixed precision results to be accurate with respect to double precision (error $\approx10^-7$) for both DNS and LES simulations. As well, in complex volcanic plume simulations, the average properties of flow are well recovered with mixed precision.

In general, the computational gain one may obtain in reduced precision varies significantly depending on the characteristics of the CFD application and the computational load per core of the test case. The latter, in particular, is directly related to how much the application is bounded by memory or MPI related communications.
Test with the lid-driven cavity in single precision shows that a speedup near to the expected $2\times$ is observed only for the test cases with a larger computational load (166k cells/core). Smaller test cases (40k cells/core) present much smaller gains (1.1-3$\times$),  due to the smaller speedup of the linear algebra part of the fluid solver. Changing the application may also affect the computational gain in reduced precision, since the fraction of time spent on different parts of the fluid solver may change significantly.  For instance, with respect to the lid-driven cavity where the linear algebra takes more than >90\% of the total time, the starting jet requires only half of the total time. 
This aspect becomes even more important when considering the computational gain one may obtain with OpenFOAM in mixed precision, for which only the matrix assembly is done in single precision and linear algebra in double precision. The lid-driven cavity solved with \texttt{icoFoam} presents indeed in general a much smaller maximum gain ($1.3\times$) than the starting jet with \texttt{rhoPimpleFoam} ($1.7\times$) or volcanic plume simulations with ASHEE.
Despite all this variability, in reduced precision (single or mixed) the matrix assembly is observed to be significantly faster for all test cases (1.7-2.2$\times$). As well, in mixed precision linear algebra is also a bit faster (1.1-1.3$\times$).

Reduced precision has also an impact on scalability. This aspect has been studied here interpreting the results of the scaling tests with the aid of an ad hoc developed theoretical model. Despite the strong assumptions, the model qualitatively reproduces the observed scaling behaviours when passing from full to reduced precision and allows us to understand and generalise the test outcomes.
At the intra-node level, the use of reduced precision improves the scalability, because the latter is affected by memory communications, as also confirmed by the modelling results and TMAM analysis. This aspect seems to be relevant for OpenFOAM (v1912) which is characterised by a low intra-node scalability performance. Moreover, the computational gain of reduced precision increases with the number of cores since the memory bandwidth per core decreases across the node. 
At the inter-node level, the scalability is more complex since both memory and MPI communications need to be considered. Supported by our scalability model results, we have shown that memory communications are mainly responsible for the superlinear behaviour observed for both double and mixed precision up to $\approx 10^4$ cells/core.  In this memory bound region, the application is indeed scaling better than in the ideal case since, as the number of nodes grows, the total cache available in the system increases, thus reducing the memory bandwidth bottleneck. The overall effect is that the application receives an additional performance boost and hence an additional contribution to the speedup. In the memory bound region of the speedup plot, mixed and double precision present similar scalability, with mixed precision performing slightly better. It is important also to note that the degree of super-linearity in a speedup plot is related to how much the reference case (in our case the one run on a single node) is bounded by memory bandwidth. Indeed, when the reference case is not strongly affected by the bandwidth bottleneck (in our case, on multiple nodes) the super-linear behaviour is not observed at all.
When MPI communications start to play a role (i.e., when the number of cells/core becomes too small), the parallel efficiency decreases quite rapidly as more computational nodes are used and the full precision even scales better than reduced precision. Nevertheless, the computational gain in reduced precision with respect to the full precision is significant, up to the number of cells/core for which the application presents good parallel scalability. 
Although the conclusions regarding the scalability are based on the comparison between double and mixed precision, we believe they can apply also for the single precision case.

Finally, reduced precision computation is also desirable when working with GPUs, given that efficient data communication is needed to properly exploit the potential of such hardware.
In the near future, all computationally intensive applications will be required to use GPUs, since they are becoming an increasingly common and fundamental part of modern heterogenous cluster architectures.  
In this work, we have demonstrated the power of using mixed precision for a hybrid CPU-GPU OpenFOAM implementation, which, at the moment, is the only option available to the end user (from the up-to-date official OpenFOAM repository). 
In the hybrid CPU-GPU approach, only the linear algebra part is offloaded on the GPU. Therefore, the total speedup achievable is primarily a function of the fraction of time spent by the application on the linear solver. The latter is application dependent and, in our tests, it may vary from 90\% for the lid-driven cavity to 50\% with compressible starting jet. Neglecting CPU-GPU communication cost and applying Amdahl's Law, it is easy to predict the total speedup for an application. When considering the starting jet (i.e., the worst-case scenario in terms of the time spent in the linear algebra solvers), the combination of mixed precision with the hybrid CPU-GPU implementation may theoretically provide up to $2.4\times$ speedup (w.r.t double precision on pure CPU architecture). This value in fact matches the one measured in our single node tests. Although preliminary, this result may represent a first indication for the end user who wants to speed up CFD simulations with hybrid CPU-GPU and mixed precision implementation.
Multi-node GPU scaling tests are left for future work. As well as more work remains to be done to test the use of mixed precision linear algebra solvers with GPUs or tensor cores that can even work in half precision \citep{Haidar2018}. Interestingly, these solvers may further speed up the resolution of linear algebra without compromising the convergence or accuracy of computed solutions. Finally, in this study, we did not consider IO and power consumption, two important aspects that we believe deserve dedicated works.

%% file: 6-appendix.tex
\clearpage

\section{Appendix}

\setcounter{figure}{0}  

\begin{figure}[ht]
\renewcommand\figurename{Figure A1}
    \centering
    \includegraphics[width=1.0\columnwidth]{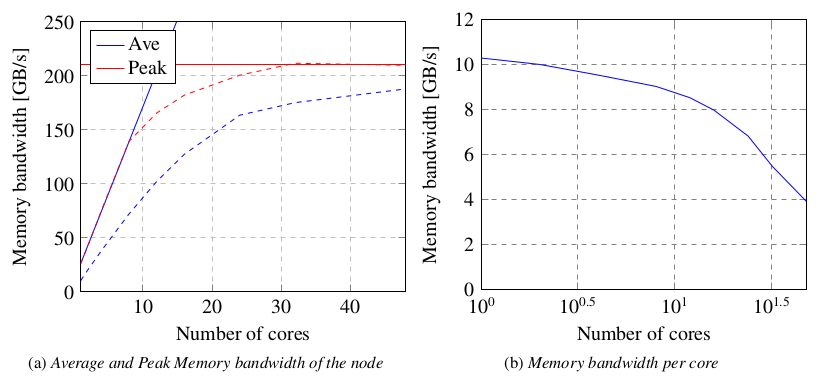}
   \caption*{Figure A1: Average and Peak Memory bandwidth registered in the MARCONI cluster using the 3D Lid-driven cavity benchmark (case M).}
\label{MemBandwidth}
\end{figure}

\subsection{Appendix A: Top-Down Microarchitecture Analysis Method (TMAM) of an OpenFAOM application}
\label{app:TMAM}
We performed Top-Down Microarchitecture Analysis Method (TMAM) of the lid-driven cavity benchmark varying the number of cores (aka strong scaling). According to the TMAM analysis \citep{TMAM14}, a slot in a CPU pipeline can be found in a binary state, stalled or not. 
Our analysis has found that Retiring and Back End bound are the two most frequent states in the first level of the TMAM. Bad speculation and Front End Bound can be neglected. At the second level, Memory Bound is the dominant state and Core Bound can be neglected. Figure A2 shows the slot fraction of Retiring and Memory Bound for the M-version of the lid-driven cavity run in double and mixed precision.
Both intra-node and inter-node tests have shown that the slot fraction of Retiring is shifted higher in the mixed precision than the double precision. As a consequence, the complementary slot fraction of Memory Bound is shifted lower.
In other words, the mixed precision increases the efficiency of the simulation but does not influence the trend with the variation of the number of cores. This trend is characterized by an increasing of the slot fraction of the memory stalls in the intra-node case due to the memory bandwidth contention. In the inter-node case, due to the increasing of the RAM and cache size of the system and the decreasing of the problem size per node, the efficiency increases until saturation, see Figure A2.

\begin{figure}[ht]
\renewcommand\figurename{Figure A2}
   \centering
   \includegraphics[width=1.0\columnwidth]{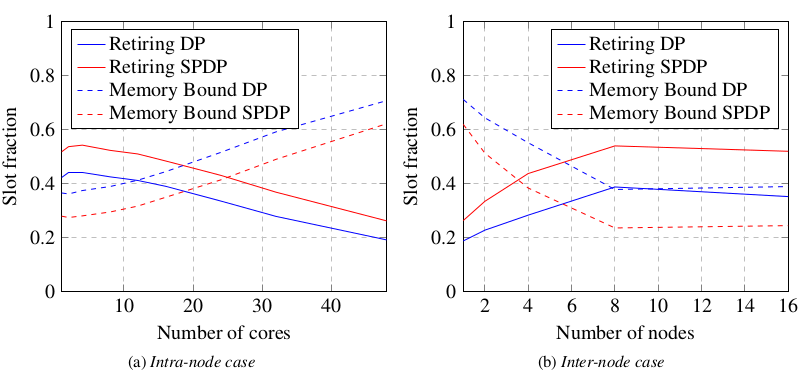}
   \caption*{Figure A2: Retiring and Memory Bound slot fraction in the TMAM analysis of the 3D lid-driven cavity benchmark (case M) using double and mixed precision.}
\end{figure}

\subsection{Appendix B: Modelling MPI communication in strong scaling experiment}
\label{app:MPI}
Our profiler analysis has shown that in double precision the most expensive and dominant MPI function is \texttt{MPI\_Allreduce} (see Figure A3). The same profiling analysis repeated with the code compiled in mixed precision provides evidence that \texttt{MPI\_Recv} is the most used.  However, we think that this result is only due to the way the mixed precision code is implemented.

In \citep{Hoefler10} the authors have shown that the behaviour of \texttt{MPI\_Allreduce} is asymptotically logarithmic. However, if the number of cores is relatively low ($P < P*$, where $P*$ is in the order of thousands), the time is constant. This behaviour has not been registered by our tests since they include the slack time due to load unbalance; 
 this term is not negligible, as confirmed by our tests, but decreases of importance as the number of cells per core decreases (see Figure A3).

\begin{figure}[ht]
\renewcommand\figurename{Figure A3}
    \centering
    \includegraphics[width=1.0\columnwidth]{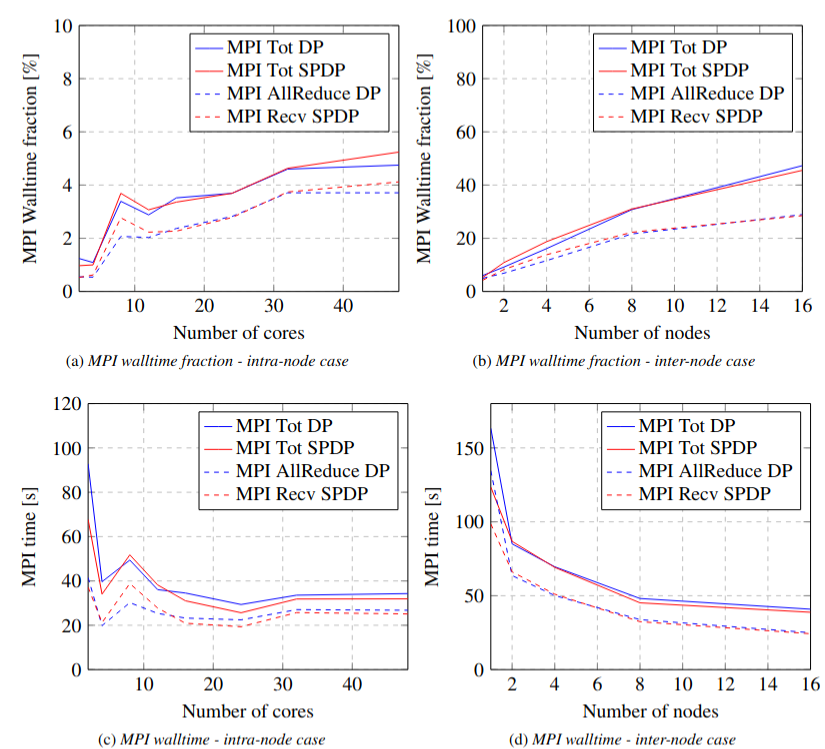}
   \caption*{Figure A3: MPI walltime fraction and absolute walltime in the strong scaling test (intra-node and inter-node) of the 3D lid-driven cavity benchmark (case M) using double and mixed precision.}
\end{figure}

%% file: 7-Aknwoledgments.tex
\clearpage

\textit{Author contributions.}{
Conceptualization F.B.; 
methodology F.B., S.B., G.B., G.A., M.C.;
simulations F.B., S.B.,  G.B., G.A., M.C.;
scaling analysis F.B., G.A., S.B.;
theoretical model S.B., G.B., F.B.;
software F.B., S.B., G.B., M.C;.
original draft preparation F.B., S.B., G.B., M.C.;
review and editing of the manuscript, all authors. 
} \newline

\textit{Competing interests.}{The Authors declare no competing interests.}\newline

\textit{Acknowledgements and Financial support.}{ This work has been supported by Istituto Nazionale di Geofisica e Vulcanologia (INGV) and by SuperCompunting Application and Innovation Department (CINECA). This research has received funding from European Union’s Horizon 2020 research and innovation programme under the ChEESE project, grant agreement no 823844 and from INGV Pianeta Dinamico grant CHOPIN.}